\documentclass[12pt]{article}

\baselineskip=\normalbaselineskip

\usepackage{mathrsfs,amsbsy,amssymb,latexsym,amsfonts,amsmath}
\usepackage{graphicx,color}

\allowdisplaybreaks[2]
\renewcommand{\thefootnote}{\arabic{footnote}}
\makeatletter
\catcode`\@=11
\@addtoreset{equation}{section}

\def\@seccntformat#1{\csname the#1\endcsname.~~}
\makeatother

\newcommand{\tr}{\operatorname{tr}}
\newcommand{\rmd}{{\mathrm d}}
\newcommand{\nn}{\nonumber}
\newcommand{\Exp}[1]{\operatorname{e}^{#1}}
\newcommand{\WGamma}[2]{\Gamma^{#1}_{\!\!{\rm (w)}\,{#2}}}
\def\cA{\mathcal A}

\def\cD{\mathcal D}
\def\cF{\mathcal F}
\def\cG{\mathcal G}
\def\cM{\mathcal M}
\def\cR{\mathcal R}

\newcommand{\bx}{{\boldsymbol{x}}}

\begin{document}

\begin{titlepage}
\renewcommand{\thefootnote}{\fnsymbol{footnote}}

\begin{flushright}
\parbox{3.5cm}
{KUNS-2391\vspace{-1mm}\\
MISC-2012-08}
\end{flushright}

\vspace*{1.0cm}

\begin{center}
{\Large \bf Conformal higher-order viscoelastic fluid mechanics}%
\end{center}
\vspace{1.0cm}

\centerline{
{Masafumi Fukuma$^1$}%
\footnote{E-mail address: 
fukuma\_at\_gauge.scphys.kyoto-u.ac.jp} ~and~ 
{Yuho Sakatani$^2$}%
\footnote{E-mail address: 
yuho\_at\_cc.kyoto-su.ac.jp}%
}

\vspace{2mm}

\begin{center}
{$^1$\it Department of Physics, Kyoto University \\ 
Kyoto 606-8502, Japan\\}
\vspace{3mm}
{$^2$\it Maskawa Institute for Science and Culture, Kyoto Sangyo University \\ 
Kyoto 603-8555, Japan\\}

\end{center}
\vspace*{1cm}
\begin{abstract}

We present a generally covariant formulation 
of conformal higher-order viscoelastic fluid mechanics 
with strain allowed to take arbitrarily large values. 
We give a general prescription to determine the dynamics 
of a relativistic viscoelastic fluid 
in a way consistent with the hypothesis of local thermodynamic equilibrium 
and the second law of thermodynamics. 
We then elaborately study the transient time scales 
at which the strain almost relaxes 
and becomes proportional to the gradients of velocity. 
We particularly show that a conformal second-order fluid 
with all possible parameters in the constitutive equations 
can be obtained without breaking the hypothesis 
of local thermodynamic equilibrium, 
if the conformal fluid is defined as the long time limit 
of a conformal second-order viscoelastic system. 
We also discuss how local thermodynamic equilibrium 
could be understood in the context of the fluid/gravity correspondence.

\end{abstract}
\thispagestyle{empty}
\end{titlepage}

\tableofcontents
\setcounter{footnote}{0}

\section{Introduction}
\label{introduction}

Viscoelasticity is the property shared by almost all continuum materials, 
showing elasticity at short time scales 
while behaving as a viscous fluid at long time scales 
\cite{LL_elasticity,Eckart:1948}. 
In papers \cite{fs1,fs2}, 
the present authors constructed relativistic viscoelastic fluid mechanics 
in a generally covariant form 
based on Onsager's linear nonequilibrium thermodynamics. 
We showed there that, 
for arbitrary parameters in the constitutive equations, 
viscoelastic fluids thus defined behave 
as standard first-order viscous fluids 
(obeying the relativistic Navier-Stokes equations \cite{Eckart:1940te,LL_fluid}) 
at long time scales \cite{fs2}. 
We also showed that the evolution equations are hyperbolic 
for a wide range of parameters 
due to the elasticity at short time scales. 
In this sense, a relativistic viscoelastic model with such parameters 
gives a causal completion of standard first-order relativistic fluid mechanics 
\cite{fs2}.

Recently, various models of relativistic fluids 
with second-order corrections in the derivative expansion 
(called second-order fluid mechanics) 
have been considered in the analysis of heavy-ion collision experiments 
and also in the study of the holographic duality between 
the long wavelength dynamics of black hole horizons 
and the dynamics of viscous fluids 
(see, e.g., \cite{Romatschke,fluid-gravity_review} and references therein). 
For both cases, fluid systems are well approximated to be invariant 
under conformal transformations. 
In particular, in \cite{Baier:2007ix} and \cite{Loganayagam:2008is}, 
a model of conformal fluid with full second-order corrections was constructed 
and shown to be consistent with the second law of thermodynamics 
for a wide range of parameters \cite{Loganayagam:2008is}.

A remarkable point for second-order fluids 
is that their entropy densities  
generally contain spatial derivatives of thermodynamic variables. 
This means that local thermodynamic equilibrium is broken for second-order fluids. 
In fact, in thermodynamics the coordinates $x=(x^0,\bx)$ 
represent a coarse-grained spacetime point, 
whose temporal and spatial resolutions we denote 
by $\epsilon_{\rm t}$ and $\epsilon_{\rm s}$\,, respectively \cite{fs1}. 
We say that local thermodynamic equilibrium is realized at $x$ 
with resolution $(\epsilon_{\rm t},\epsilon_{\rm s})$ 
if a small spatial region around $\bx$ (of linear size $\epsilon_{\rm s}$) 
at time $x^0$ 
can be well regarded as being in thermodynamic equilibrium 
at least for a time duration of $\epsilon_{\rm t}$\,.%
\footnote{
We call such spatial regions {\em material particles}. 
} 
This implies that the local entropy in the coarse-grained region around $x$ 
is already maximized for given values of local thermodynamic variables at $x$ 
(such as the energy-momentum density and the charge density), 
and thus that the entropy density at $x$\,, $s(x)$,  
is a function only of these local thermodynamic variables 
and should not depend on their spatial derivatives 
(which include contributions from nearby material particles). 
Thus, when we consider thermodynamics of second-order fluids, 
we are generally forced to give up the hypothesis of local thermodynamic equilibrium 
and to use (a variant of) {\em extended thermodynamics}, 
where the entropy density can depend on quantities including spatial derivatives 
(such as dissipative currents) \cite{Mueller,Israel:1976tn,Israel:1979wp}.

In this paper, introducing the elastic strain tensor 
as one of local thermodynamic variables, 
and utilizing the manifestly Weyl-covariant formulation 
of conformal second-order fluid mechanics 
developed by Loganayagam \cite{Loganayagam:2008is}, 
we construct conformal higher-order (i.e., not first-order) 
{\em viscoelastic} fluid mechanics in the Landau-Lifshitz frame 
in such a way that local thermodynamic equilibrium 
and the second law of thermodynamics are manifestly realized.  
We show that the conformal second-order fluid mechanics of 
\cite{Baier:2007ix,Loganayagam:2008is} 
with all possible parameters in the constitutive equations
can be fully recovered as the long time limit of our viscoelastic model 
of second order. 
Thus, if we define conformal second-order fluid mechanics 
as the long time limit of a conformal viscoelastic system,  
conformal second-order fluid mechanics can be constructed 
without violating the hypothesis of local thermodynamic equilibrium.%
\footnote{
We comment that local thermodynamic equilibrium 
can also be realized in divergence-type fluid mechanics 
developed by Geroch and Lindblom \cite{Geroch:1990bw,PeraltaRamos:2009kg}. 
There, a symmetric traceless tensor $\xi_{\mu\nu}$ 
is introduced as a local dynamical variable 
in addition to the standard dynamical variables 
$\xi\equiv \mu/T$ and $\xi_\mu\equiv u_\mu/T$ 
($T$: temperature, $u^\mu$: velocity, $\mu$: chemical potential). 
In fact, the causally completed extensions of Eckart's fluid mechanics \cite{Geroch:1990bw} 
and of Landau-Lifshitz's \cite{PeraltaRamos:2009kg} satisfy local thermodynamic equilibrium 
in the sense that the entropy density depends only on 
$\xi$\,, $\xi_\mu$\, and $\xi_{\mu\nu}$\,. 
The apparent difference between divergence-type fluid mechanics 
and viscoelastic fluid mechanics 
is that the additional dynamical variable in the former 
is related to the conserved current (often denoted by $A^{\lambda\mu\nu}$), 
while the strain tensor introduced in the latter 
has no origin as a conserved quantity. 
A possible relationship between two theories will be commented in section \ref{transient}. 
} 

This paper is organized as follows. 
In section 2, we briefly review a part of the manifestly Weyl-covariant 
formulation of conformal second-order fluid mechanics \cite{Loganayagam:2008is}.  
In section 3, assuming local thermodynamic equilibrium, 
we present a general theory describing relativistic viscoelastic fluids with large strain. 
In section 4, we construct conformal higher-order viscoelastic fluid mechanics, 
based on the manifestly Weyl-covariant formalism. 
In section 5, we investigate the transient time scales 
at which the strain almost relaxes and becomes 
proportional to the gradients of velocity. 
We there verify our claim that a conformal second-order fluid 
with all possible parameters in the constitutive equations 
can be obtained as the long time limit of a conformal viscoelastic system 
that satisfies the hypothesis of local thermodynamic equilibrium.  
In section 6, we briefly discuss how local thermodynamic equilibrium 
could be understood in the context of the fluid/gravity correspondence, 
and point out that its manifest realization may lead to 
the viscoelasticity/quantum gravity correspondence. 
Appendix \ref{weight_list} gives the list of 
the dimensions and the weights of various local thermodynamic quantities, 
and Appendix \ref{formulas} collects useful formulas 
in the manifestly Weyl-covariant formalism with proofs.

\section{Manifestly Weyl-covariant formalism}
\label{Weyl}

In this section, in order to fix our notation, 
we give a brief review on the manifestly Weyl-covariant formulation 
of second-order fluid mechanics developed in \cite{Loganayagam:2008is} 
(see also \cite{Bhattacharyya:2008xc,Bhattacharyya:2008mz}).

We consider a $d$-dimensional spacetime with background metric $g_{\mu\nu}$ 
of signature $(-,+,\allowbreak \ldots,\allowbreak +)$. 
A tensor $Q^{\mu\ldots}_{\nu\ldots}$ is called a conformal tensor of weight $w$ 
if it transforms as
\begin{align}
 Q^{\mu\ldots}_{\nu\ldots}= \Exp{w\phi}\, Q^{\,\prime\mu\ldots}_{\,\,\nu\ldots} 
\end{align}
under the Weyl transformation%
\footnote{
Note that $g_{\mu\nu}$ itself is a conformal tensor of weight $w=-2$\,. 
The dimensions and the weights of various local thermodynamic quantities 
are listed in Appendix \ref{weight_list}. 
}
\begin{align}
 g_{\mu\nu} = \Exp{-2\phi} \, g^{\,\prime}_{\mu\nu} \,.
\label{Weyl_transformation}
\end{align}
We introduce the Weyl connection 
$\WGamma{\lambda}{\mu\nu}\equiv \Gamma^{\lambda}_{~\mu\nu} 
  + W^{\lambda}_{~\mu\nu}$ with
\begin{align}
 \Gamma^{\lambda}_{~\mu\nu} \equiv \frac{1}{2}\, g^{\lambda\sigma}\,
  \bigl(\partial_\mu g_{\nu\sigma} + \partial_\nu g_{\mu\sigma}
  - \partial_\sigma g_{\mu\nu}\bigr)\,, \quad
  W^{\lambda}_{~\mu\nu} \equiv 
  g_{\mu\nu}\,\cA^\lambda - \delta^\lambda_\mu\, \cA_{\nu}
  - \delta^\lambda_\nu\, \cA_\mu \,.
\end{align}
$\WGamma{\lambda}{\mu\nu}$ is invariant under the Weyl transformation 
\eqref{Weyl_transformation} 
if the (non-conformal) vector field $\cA_\mu$ transforms as
\begin{align}
 \cA_{\mu} = \cA^{\,\prime}_{\mu} - \partial_\mu \phi \,. 
\label{Weyl_transformation_A}
\end{align}
Following \cite{Loganayagam:2008is}, 
we introduce Dirac's {\em co-covariant derivative} \cite{Dirac:1973gk} 
for a conformal tensor $Q^{\mu\ldots}_{\nu\ldots}$ of weight $w$ as
\begin{align}
 \cD_\lambda Q^{\mu\ldots}_{\nu\ldots} 
 &\equiv \partial_\lambda Q^{\mu\ldots}_{\nu\ldots} 
  + \WGamma{\mu}{\lambda\rho}\,Q^{\rho\ldots}_{\nu\ldots}+\cdots
  - \WGamma{\rho}{\lambda\nu}\,Q^{\mu\ldots}_{\rho\ldots}-\cdots
  + w\,\cA_\lambda\,Q^{\mu\ldots}_{\nu\ldots} 
\nn\\
 &= \nabla_\lambda Q^{\mu\ldots}_{\nu\ldots} 
  + W^{\mu}_{~\lambda\rho}\,Q^{\rho\ldots}_{\nu\ldots}+\cdots
  - W^{\rho}_{~\lambda\nu}\,Q^{\mu\ldots}_{\rho\ldots}-\cdots
  + w\,\cA_\lambda\,Q^{\mu\ldots}_{\nu\ldots} \,,
\end{align}
which enjoys the following properties \cite{Loganayagam:2008is}:
\begin{align}
 \cD_\lambda Q^{\mu\ldots}_{\nu\ldots}
  &= \Exp{w\phi}\, \cD^{\,\prime}_\lambda 
  Q^{\,\prime\mu\ldots}_{\,\nu\ldots} 
  \quad \mbox{if} \quad 
  Q^{\mu\ldots}_{\nu\ldots}= \Exp{w\phi}\, 
  Q^{\,\prime\mu\ldots}_{\,\nu\ldots} \,,
\\
 \cD_\lambda g_{\mu\nu} &=0 \,, \qquad \cD_\lambda g^{\mu\nu} =0 \,.
\end{align}
Note that if a contravariant vector $v^\mu$ has weight $w=d$, 
the following equality holds:
\begin{align}
 \cD_\mu v^\mu = \nabla_\mu v^\mu + (w-d)\,\cA_\mu v^\mu = \nabla_\mu v^\mu
 \quad \mbox{(when $w=d$)}\,.
\label{divergence_vector}
\end{align}
Similarly, if a $(2,0)$ tensor $Q^{\mu\nu}$ is symmetric traceless 
and has weight $w=d+2$\,, 
we have  
\begin{align}
 \cD_\mu\,Q^{\mu\nu}=\nabla_\mu\,Q^{\mu\nu}
  \quad \mbox{(when $w=d+2$\,, $Q^{\mu\nu}=Q^{\nu\mu}$ and $Q^\mu_\mu=0$)}\,.
\label{divergence_tensor}
\end{align}

Given the velocity field $u^\mu$ for a fluid 
(having unit weight and being normalized as $u^\mu\,u_\mu=-1$), 
the vector field $\cA_\mu$ can be uniquely determined 
by requiring the covariant derivative of $u^{\mu}$
to be transverse ($u^\nu\, \cD_\nu u^\mu =0$) and 
divergenceless ($\cD_\mu u^\mu =0$) \cite{Loganayagam:2008is}:
\begin{align}
 \cA^{\mu} = a^\mu - \frac{\vartheta}{d-1} \, u^{\mu} \,,
\label{cA_def_a_mu}
\end{align}
where $a^\mu$ and $\vartheta$ are the acceleration and the expansion, 
respectively:
\begin{align}
 a^\mu &\equiv \nabla_u u^\mu \equiv u^\nu\,\nabla_\nu u^\mu\,,\qquad
  \vartheta \equiv \nabla_\mu u^\mu \,.
\end{align}
In this paper, 
we write the (anti-)symmetrization of indices 
as $Q_{(\mu\nu)}\equiv(1/2)\bigl(Q_{\mu\nu}+ Q_{\nu\mu}\bigr)$ 
and $Q_{[\mu\nu]}\equiv(1/2)\bigl(Q_{\mu\nu}- Q_{\nu\mu}\bigr)$. 
We further denote the symmetric, transverse, traceless part 
of $Q_{\mu\nu}$ by 
\begin{align}
 Q_{\langle\mu\nu\rangle}\equiv
  h_{\mu}^\alpha h_{\nu}^\beta Q_{(\alpha\beta)}
  -\frac{1}{d-1}\,(h^{\alpha\beta} Q_{\alpha\beta})\,h_{\mu\nu}\,,
\end{align}
where $h_{\mu\nu}\equiv g_{\mu\nu}+u_\mu u_\nu$ is 
the metric projected to a surface orthogonal to $u^\mu$\,.
One then can show \cite{Loganayagam:2008is} 
that $\cD_\mu u^\nu$ can be decomposed as
\begin{align}
 \cD_{\mu}u^\nu 
  &=\sigma_{\mu}^{~\nu} + \omega_{\mu}^{~\nu} 
\end{align}
with%
\footnote{
Note that $\sigma_{\mu\nu}=\sigma_{\langle\mu\nu\rangle}$\,. 
We will use the abbreviation such as 
$(\sigma^2)_{\mu\nu}\equiv\sigma_{\mu}^{~\alpha}\,\sigma_{\alpha\nu}$ 
and $\tr (\sigma^2)\equiv \sigma_{\alpha}^{~\beta}\sigma_{\beta}^{~\alpha}$\,. 
It then holds that 
$(\sigma^n)_{\mu\nu}=(\sigma^n)_{(\mu\nu)}
=h_{\mu}^{~\alpha}h_{\nu}^{~\beta}(\sigma^n)_{(\alpha\beta)}$ 
for $n=1,2,\ldots,$ 
and 
$(\omega^n)_{\mu\nu}$ is $h_{\mu}^{~\alpha}h_{\nu}^{~\beta}
(\omega^n)_{(\alpha\beta)}$  for even $n$ 
and $h_{\mu}^{~\alpha}h_{\nu}^{~\beta}\,(\omega^n)_{[\alpha\beta]}$ 
for odd $n$\,.  
}
\begin{align}
 \sigma_{\mu\nu}&\equiv \cD_{(\mu}u_{\nu)}
  =\nabla_{\langle\mu}u_{\nu\rangle}\,,\qquad
  \omega_{\mu\nu}\equiv \cD_{[\mu} u_{\nu]}
  =h_\mu^\rho\,h_\nu^\sigma\,\nabla_{[\rho}u_{\sigma]}\,.
\end{align}

We introduce the Weyl-covariantized Riemann tensor 
$\cR_{\mu\nu\lambda}{}^{\sigma}$ \cite{Loganayagam:2008is} 
as the curvature for the Weyl connection $\WGamma{\lambda}{\mu\nu}$:%
\footnote{
We follow the convention of \cite{Bhattacharyya:2008xc,Bhattacharyya:2008mz} 
where all the curvature tensors are negative of those in \cite{Loganayagam:2008is}. 
Tensors of non-calligraphic font are the ones 
constructed from the affine connection $\Gamma^\lambda_{\mu\nu}$ 
(i.e.\ the ones obtained by replacing $\cD_\mu$ by $\nabla_\mu$).
}
\begin{align}
 \cR_{\mu\nu\lambda}{}^{\sigma}\equiv
  -\,\partial_\mu\WGamma{\sigma}{\nu\lambda}+\partial_\nu\WGamma{\sigma}{\mu\lambda}
  -\WGamma{\sigma}{\mu\rho}\,\WGamma{\rho}{\nu\lambda}
  +\WGamma{\sigma}{\nu\rho}\,\WGamma{\rho}{\mu\lambda}
  = \cR^{\prime}_{\mu\nu\lambda}{}^{\sigma}\,.
\label{Riemann_tensor_conformal}
\end{align}
The explicit form is given by 
\begin{align}
 \cR_{\mu\nu\lambda\sigma} &= R_{\mu\nu\lambda\sigma} 
   + \cF_{\mu\nu} \, g_{\lambda\sigma}
   - 4\,\delta^\alpha_{[\mu}g_{\nu][\lambda}\delta^\beta_{\sigma]}
       \,\Bigl(\nabla_\alpha \cA_\beta + \cA_\alpha \cA_\beta
   - \frac{\cA^2}{2} g_{\alpha\beta} \Bigr) 
  = \Exp{-2\phi}\,\cR^{\prime}_{\mu\nu\lambda\sigma}\,.
\label{Weyl_Riemann_A_F}
\end{align}
It is easy to see that the following equality holds 
for a covariant vector $v_\lambda$ of weight $w$ \cite{Loganayagam:2008is}:\begin{align}
 [\cD_\mu,\cD_\nu]\, v_\lambda 
  &= \cR_{\mu\nu\lambda}{}^{\sigma}\, v_\sigma + w\, \cF_{\mu\nu}\, v_\lambda \,,
\\
 \cF_{\mu\nu} &\equiv \partial_\mu\cA_\nu - \partial_\nu\cA_\mu \,.
\label{A_F}
\end{align}
For a contravariant vector $v_1^{\lambda}$ of weight $w_1$, we obtain%
\footnote{
Note that $w_1=w+2$ if $v_1^{\lambda}=g^{\lambda\mu}\,v_\mu$\,.
}
\begin{align}
 [\cD_\mu,\cD_\nu]\, v_1^{\lambda} 
  &=- \cR_{\mu\nu\sigma}{}^{\lambda}\, v_1^{\sigma}
  + w_1\, \cF_{\mu\nu}\, v_1^{\lambda}\,.
\end{align}
We also introduce the following conformal tensors:
\begin{align}
 \cR_{\mu\nu}&\equiv  \cR_{\mu\alpha\nu}{}^{\alpha} 
  =  R_{\mu\nu} + \cF_{\mu\nu}
   + \nabla_\rho\cA^\rho \, g_{\mu\nu}
   + (d-2)\,\bigl(\nabla_{\mu} \cA_{\nu} 
   + \cA_\mu\, \cA_\nu -\cA^2 g_{\mu\nu} \bigr)
  = \cR^{\,\prime}_{\mu\nu}\,,
\label{Weyl_Ricci_A_F}
\\
 \cR &\equiv \cR_{\alpha}^{~\alpha}
  = R + 2(d-1)\,\nabla_\rho \cA^\rho - (d-1)\,(d-2)\, \cA^2
  = \Exp{2\phi}\,\cR^{\prime} \,,\\
 \cG_{\mu\nu} 
  &\equiv \cR_{\mu\nu}-\frac{1}{2}\, \cR\,g_{\mu\nu}\,, \nn\\
  &= G_{\mu\nu} + \cF_{\mu\nu}
   + (d-2)\,\Bigl[\nabla_{\mu} \cA_{\nu} + \cA_\mu\, \cA_\nu 
   - \Bigl(\nabla_\rho\cA^\rho -\frac{d-3}{2}\,\cA^2\Bigr)\,g_{\mu\nu} \Bigr]
   =\cG^{\,\prime}_{\mu\nu}\,.
\label{Weyl_R_A_F}
\end{align}
These tensors have the following symmetry properties for their indices 
\cite{Loganayagam:2008is}: 
\begin{align}
 \cR_{(\mu\nu)\,\lambda\sigma}&=0\,,\qquad
  \cR_{\mu\nu\,(\lambda\sigma)}=\cF_{\mu\nu}\,g_{\lambda\sigma}\,,\qquad
  \cR_{[\lambda\mu\nu]\sigma}=0\,,
\label{indices1}
\\
 \cR_{\mu\nu\,\lambda\sigma}&=\cR_{\lambda\sigma\,\mu\nu}
  + 4\,\delta^\alpha_{[\mu}g_{\nu][\lambda}\delta^\beta_{\sigma]}\,\cF_{\alpha\beta}
  + \cF_{\mu\nu}\,g_{\lambda\sigma} - g_{\mu\nu}\,\cF_{\lambda\sigma}\,,
\label{indicies2}
\\
 v^\alpha\,\cR_{\alpha\,[\mu\nu]\,\beta}\,v^\beta &= 
  -\,\frac{1}{2}\,\cF_{\mu\nu}\,v^2\,,\qquad
  \cR_{[\mu\nu]}=\frac{d}{2}\,\cF_{\mu\nu}\,.
\label{indices3}
\end{align}
The Bianchi identity $\bigl(\,[\cD_\lambda,\,[\cD_\mu,\,\cD_\nu]\,]
+\mbox{(cyclic)}\,\bigr)\,v_\rho=0$ gives \cite{Loganayagam:2008is}
\begin{align}
 \cD_\lambda \cR_{\mu\nu\rho}{}^{\sigma}
  +\cD_\mu \cR_{\nu\lambda\rho}{}^{\sigma}
  +\cD_\nu \cR_{\lambda\mu\rho}{}^{\sigma}&=0\,,\\
 \nabla_\lambda \cF_{\mu\nu}+\nabla_\mu \cF_{\nu\lambda}
  +\nabla_\nu \cF_{\lambda\mu} &= 0\,.
\end{align}
Contracting the indices in the first equation, we obtain
\begin{align}
 &\cD_\alpha  \cR_{\mu\nu\lambda}{}^{\alpha}
  +\cD_\mu\cR_{\nu\lambda}-\cD_\nu\cR_{\mu\lambda} = 0\,,\\
 &\cD_\nu\cR^{\mu\nu}=\frac{1}{2}\,\cD^\mu\cR + \cD_\nu\cF^{\mu\nu} \,.
\end{align}
The last equation is equivalent to
\begin{align}
 0=\cD_\mu\bigl(\cG^{\nu\mu}-\cF^{\nu\mu}\bigr)
  =\cD_\mu\Bigl( \cG^{(\mu\nu)}-\frac{d-2}{2}\,\cF^{\mu\nu}\Bigr)\,.
\label{DGF}
\end{align}

The Weyl tensor is defined as
\begin{align}
 C_{\mu\nu\lambda\sigma}\equiv R_{\mu\nu\lambda\sigma}
  +\frac{4}{d-2}\,\,\delta^\alpha_{[\mu}g_{\nu][\lambda}\delta^\beta_{\sigma]}\,
  \Bigl( R_{\alpha\beta}-\frac{R}{2(d-1)}\,g_{\alpha\beta}\Bigr)
  = \Exp{-2\phi}\,C^{\,\prime}_{\mu\nu\lambda\sigma}\,.
\label{Weyl_tensor}
\end{align}
Although this is given only in terms of the metric, 
we can rewrite it as a sum of Weyl-covariantized curvature tensors 
[see Eq.~\eqref{Weyl_tensor_conformal2}]: 
\begin{align}
 C_{\mu\nu\lambda\sigma} = \cR_{\mu\nu\lambda\sigma}
  -\cF_{\mu\nu}\,g_{\lambda\sigma}
  +\frac{4}{d-2}\,\,\delta^\alpha_{[\mu}g_{\nu][\lambda}\delta^\beta_{\sigma]}\,
  \Bigl( \cR_{\alpha\beta}-\frac{\cR}{2(d-1)}\,g_{\alpha\beta}
  +\cF_{\alpha\beta}\Bigr)\,.
\label{Weyl_tensor_conformal}
\end{align}
When $\cA_\mu$ is constructed from the conformal velocity 
field $u^\mu$ of unit weight as in Eq.\ \eqref{cA_def_a_mu}, 
we have the following formulas 
[see Eqs.~\eqref{uCu_uRu2} and \eqref{u_cR_u_trace}]:
\begin{align}
 u^\alpha\,\cR_{\alpha\,\langle\mu\nu\rangle\,\beta}\,u^\beta
  &=u^\alpha\,C_{\alpha\mu\nu\beta}\,u^\beta \,
  +\frac{1}{d-2}\,\cR_{\langle\mu\nu\rangle}\,,
\label{uCu_uRu}
\\
 u^\alpha\,\cR_{\alpha\beta}\,u^\beta &= -\tr(\sigma^2)-\tr(\omega^2)\,.
\end{align}
The shear $\sigma_{\mu\nu}$ and the vorticity $\omega_{\mu\nu}$ 
satisfy the following equalities 
$\bigl(\cD_u\equiv u^\mu\cD_\mu\bigr)$ 
[see Eqs.~\eqref{DDsigma}, \eqref{DDomega}, 
\eqref{u_cR_u_symm}, \eqref{uCu_uRu2} and \eqref{u_cR_u_anti}]:
\begin{align}
 \cD_\mu\cD_\nu\sigma^{\mu\nu} &= \cR^{\langle\mu\nu\rangle}\,\sigma_{\mu\nu}
  - \frac{1}{2}\,\cF^{\mu\nu}\omega_{\mu\nu}+\frac{1}{2}\,\cD_u\,\cR
  + (d-2)\,\cD_\mu(\cF^{\mu\nu}\,u_\nu)\,, \\
 \cD_\mu\cD_\nu\omega^{\mu\nu}
  &= -\,\frac{d-3}{2}\,\cF^{\mu\nu}\omega_{\mu\nu}\,,\\
 \cD_u\sigma_{\mu\nu} &= u^\alpha\,C_{\alpha\mu\nu\beta}\,u^\beta
  +\frac{1}{d-2}\,\cR_{\langle\mu\nu\rangle}
  -(\sigma^2)_{\langle\mu\nu\rangle} - (\omega^2)_{\langle\mu\nu\rangle}\,,
\label{dsigma_uCu}
\\
 \cD_u\omega_{\mu\nu} &=
  \frac{1}{2}\,h_{\mu}^{~\lambda}\,h_{\nu}^{~\sigma}\,\cF_{\lambda\sigma}
  -(\sigma\omega+\omega\sigma)_{\mu\nu}\,.
\end{align}
One can further show that 
\begin{align}
 \cD_\mu\bigl(u_\nu\cR^{\nu\mu}\bigr)
  &= \cR^{\mu\nu}\sigma_{\mu\nu}+\frac{1}{2}\,\cD_u\cR
  -\frac{d-2}{2}\,\cF^{\mu\nu}\omega_{\mu\nu}-\cD_\mu(\cF^{\mu\nu}u_\nu)\nn\\
 &=\cD_\mu\cD_\nu\sigma^{\mu\nu} -\frac{d-3}{2}\,\cF^{\mu\nu}\omega_{\mu\nu}
  -(d-1)\,\cD_\mu(\cF^{\mu\nu}u_\nu)\,,
\\
 \cD_\mu\bigl(u_\nu\cR^{\nu\rho} h_{\rho}^{~\mu}\bigr)
  &= 2\tr(\sigma^3)+6\tr(\sigma\omega^2)
  -2\sigma^{\mu\nu}\,u^\alpha\,C_{\alpha\mu\nu\beta}\,u^\beta
  -\frac{2}{d-2}\,\cR^{\mu\nu}\sigma_{\mu\nu}\nn\\
 &~~-\,\frac{d-5}{2}\,\cF^{\mu\nu}\omega_{\mu\nu}
  +\cD_\mu\cD_\nu\sigma^{\mu\nu} - (d-1)\cD_\mu(\cF^{\mu\nu}u_\nu)\,.
\label{D_uRR}
\end{align}

\section{Relativistic viscoelastic fluids with large strain}
\label{large_strain}

In this paper, we consider a conformal viscoelastic fluid 
in a $d$-dimensional spacetime with background metric $g_{\mu\nu}(x)$\,. 
We assume in this section that there exists a conserved charge 
(such as particle number) in addition to energy and momentum, 
and later will ignore the charge 
(or set the corresponding chemical potential to be zero)
to simplify discussions and expressions. 
A viscoelastic fluid then has the following local thermodynamic variables 
\cite{fs1,fs2}:
\begin{align}
 p_\mu(x)\,,\qquad n(x)\,,\qquad g_{\mu\nu}(x)\,,\qquad  \varepsilon_{\mu\nu}(x)\,. 
\end{align}
Here, $p_\mu(x)$ is the energy-momentum vector, 
$n(x)$ the charge density, 
and $\varepsilon(x)=\bigl(\varepsilon_{\mu\nu}(x)\bigr)$ the strain tensor, 
which is assumed to be spatial ($\varepsilon_{\mu\nu}\,u^\nu=0$) 
and symmetric ($\varepsilon_{\mu\nu}=\varepsilon_{\nu\mu}$) \cite{fs1,fs2}. 
We define the proper energy density as
\begin{align}
 e(x)\equiv \sqrt{-g^{\mu\nu}(x)\,p_\mu(x)\,p_\nu(x)} \,,
\end{align}
which is essentially the sum of the rest mass energy density 
and the internal energy density. 
Note that $p_\mu$ are always additive quantities, 
but $e$ is not in a relativistic theory.

In this paper, we work in the Landau-Lifshitz frame, 
where the velocity field $u=u^\mu\,\partial_\mu$ is defined 
as $u^\mu\equiv p^\mu/e$ (note that $u^\mu\,u_\mu=-1$).%
\footnote{
In this paper, we lower (or raise) indices always with $g_{\mu\nu}$
(or with its inverse $g^{\mu\nu}$). 
} 
The metric projected to a surface orthogonal to $u^\mu$ is defined as
\begin{align}
 h_{\mu\nu}\equiv g_{\mu\nu}+u_\mu u_\nu\,, 
\end{align}
and represents the shape of a material \cite{LL_elasticity,fs2}. 
By writing the velocity field as%
\footnote{
We denote spacetime coordinates by $x=(x^\mu)=(x^0,x^i)$ 
($\mu=0,1,\ldots,d-1;\,i=1,\ldots,d-1$).
Note that we need not assume $u^\mu$ to be hypersurface orthogonal 
when considering only infinitesimal neighbors around a spacetime point. 
} 
\begin{align}
 u=u^\mu\,\partial_\mu = \frac{1}{N}\,\partial_0 
  + \frac{N^i}{N}\,\partial_i
\label{ADM1}
\end{align}
with the lapse function $N$ and the shift functions $N^i$\,, 
the metric $g_{\mu\nu}$ can be expressed 
with the following ADM parametrization:
\begin{align}
 \rmd s^2&=-N^2\,\bigl(\rmd x^0\bigr)^2
  + h_{ij}\,\bigl(\rmd x^i-N^i\,\rmd x^0\bigr)\,\bigl(\rmd x^j-N^j\,\rmd x^0\bigr)\,.
\label{ADM2}
\end{align}
Note that $h_{\mu\nu}$ and $h^{\mu\nu}$ are written as
\begin{align}
 h_{\mu\nu}&=
  \begin{pmatrix}
    h_{00} & h_{0j} \\
    h_{i0} & h_{ij}
  \end{pmatrix}  
  =
  \begin{pmatrix}
    h_{kl}\,N^k N^l & -\,h_{jk}\,N^k \\
    -\,h_{ik}\,N^k & h_{ij}
  \end{pmatrix}\,,
\quad
 h^{\mu\nu}=
  \begin{pmatrix}
    0 & 0 \\
    0 & \bigl(h^{-1}\bigr)^{ij}
  \end{pmatrix}\,,
\end{align}
where $h^{-1}$ is the inverse matrix of 
the $(d-1)\times(d-1)$ matrix $h=(h_{ij})$\,. 
Thus, the local volume 
$\sqrt{h}\equiv \sqrt{\det(h_{ij})}$
satisfies the identity%
\footnote{
To obtain the last equality, 
we have used the identity 
$h^{\mu\nu}\,\delta\bigl(u_\mu u_\nu\bigr)=0$\,.
} 
\begin{align}
 \delta\sqrt{h}=\frac{\sqrt{h}}{2}\,\bigl(h^{-1}\bigr)^{ij}\,\delta h_{ij}
  =\frac{\sqrt{h}}{2}\,h^{\mu\nu}\,\delta h_{\mu\nu}
  =\frac{\sqrt{h}}{2}\,h^{\mu\nu}\,\delta g_{\mu\nu}\,.
\label{local_volume}
\end{align}

We assume that local thermodynamic equilibrium is realized 
at each spacetime point, 
so that the local entropy density $s(x)$ at spacetime point $x$ is given 
as a function of the values of these local thermodynamic variables 
at the same point $x$, 
$s(x)\allowbreak =\allowbreak s\bigl(p_\mu(x),\,\allowbreak 
n(x),\,\allowbreak g_{\mu\nu}(x),\allowbreak \varepsilon_{\mu\nu}(x)\bigr)$.  
In order to treat conserved quantities, 
it is often convenient to multiply densities by the local volume $\sqrt{h}$\,,
and we write such multiplied densities 
with placing tilde on the original densities. 
Then, the fundamental relation for the local entropy $\tilde{s}=\sqrt{h}\,s$ 
is given by \cite{fs1,fs2}
\begin{align}
 \delta \tilde{s}
  =-\,\frac{u^\mu}{T}\,\delta \tilde{p}_\mu-\frac{\mu}{T}\,\delta\tilde{n}
  +\frac{\sqrt{h}}{2T}\,T_0^{\mu\nu}\,\delta g_{\mu\nu}
  +\frac{\sqrt{h}}{T}\,S^\mu_\nu\,\delta\varepsilon_\mu^\nu\,.
\label{fundamental1}
\end{align}
Here, $T$ and $\mu$ are the temperature and the chemical potential, respectively, 
and $T_0^{\mu\nu}=e\,u^\mu u^\nu+\tau_0^{\mu\nu}$ 
is the quasi-conservative energy-momentum tensor 
with $\tau_0^{\mu\nu}$ the quasi-conservative stress tensor 
($\tau_0^{\mu\nu}=\tau_0^{\nu\mu}$, $\tau_0^{\mu\nu}\,u_\nu=0$).%
\footnote{
$T_0^{\mu\nu}$ and $\tau_0^{\mu\nu}$ were 
denoted by $T_{\rm (q)}^{\mu\nu}$ and $\tau_{\rm (q)}^{\mu\nu}$\,, 
respectively, in \cite{fs1,fs2}.
} 
We write the above with $\varepsilon_\mu^\nu$ 
(not with $\varepsilon_{\mu\nu}$) for later convenience. 
$S^\mu_\nu$ is assumed to be spatial and symmetric ($S^{\mu\nu}=S^{\nu\mu}$) 
and represents the entropic (not energetic) force caused by strain.%
\footnote{
We call $S^\mu_\nu$ the entropic force 
since it works such as to maximize the local entropy $\tilde{s}$\,. 
This is in contrast to the (energetic) 
elastic force that is exerted on a material particle 
through the spatial divergence of the stress tensor 
(see Eq.~\eqref{cons_trans}). 
} 
If we rewrite $\delta\tilde{p}_\mu$ and $\delta g_{\mu\nu}$ 
in Eq.~\eqref{fundamental1} 
in terms of $\delta\tilde{e}$\,, $\delta u_\mu$ and $\delta h_{\mu\nu}$ 
by using $\tilde{p}_\mu=\tilde{e}\,u_\mu$ 
and $g_{\mu\nu}=h_{\mu\nu}-u_\mu\,u_\nu$\,, 
then terms proportional to $\delta u_\mu$ are all canceled, 
and we obtain another expression for the fundamental relation:
\begin{align}
 \delta \tilde{s}
  = \frac{1}{T}\,\delta \tilde{e}-\frac{\mu}{T}\,\delta\tilde{n}
  +\frac{\sqrt{h}}{2T}\,\tau_0^{\mu\nu}\,\delta h_{\mu\nu}
  +\frac{\sqrt{h}}{T}\,S^\mu_\nu\,\delta\varepsilon_\mu^\nu\,.
\label{fundamental2}
\end{align}

We assume the densities with tilde to be extensive:%
\footnote{
If we use instead $\varepsilon_{\mu\nu}$ as an independent variable, 
then it should be scaled in the same way as that for $h_{\mu\nu}$\,; 
$\varepsilon_{\mu\nu}\to\lambda^{2/(d-1)}\,\varepsilon_{\mu\nu}$\,.
}
\begin{align}
 \lambda\,\tilde{s}\bigl(\tilde{e},\,\tilde{n},\,
  h_{\mu\nu},\,\varepsilon_\mu^\nu\bigr)
  =\tilde{s}\bigl(\lambda\,\tilde{e},\,\lambda\,\tilde{n},\,
  \lambda^{2/(d-1)}\,h_{\mu\nu},\,
  \varepsilon_\mu^\nu\bigr) \quad(\forall\lambda>0)\,.
\end{align}
Then, taking a derivative with respect to $\lambda$ 
and setting $\lambda=1$ afterwards, 
we obtain the following Euler relation,
having the same form as that for a fluid:
\begin{align}
 \tilde{s}=\frac{ \tilde{e}+\tilde{P}-\mu\,\tilde{n}}{T} \quad
  \quad \mbox{or~}\quad s=\frac{e+P-\mu\, n}{T}\,. 
\label{Euler}
\end{align}
Here, $P\equiv (\tr \tau_0) / (d-1)\equiv g_{\mu\nu}\,\tau_0^{\mu\nu}/(d-1)
=h_{\mu\nu}\,\tau_0^{\mu\nu}/ (d-1)$  
corresponds to the pressure of an isotropic fluid. 
From Eqs.~\eqref{local_volume}, \eqref{fundamental2} and \eqref{Euler}, 
we obtain the fundamental relation for $s$\,,
\begin{align}
 \delta s&= \frac{1}{T}\,\delta e-\frac{\mu}{T}\,\delta n
  +\frac{1}{2T}\,\tau_0^{\langle\mu\nu\rangle}\,\delta h_{\mu\nu}
  +\frac{1}{T}\,S^\mu_\nu\,\delta\varepsilon_\mu^\nu\,, 
\label{fundamental3}
\end{align}
together with the Gibbs-Duhem relation,
\begin{align}
 0&=e\,\delta\Bigl(\frac{1}{T}\Bigr) + n\,\delta\Bigl(-\frac{\mu}{T}\Bigr)
  +\delta\Bigl(\frac{P}{T}\Bigr)
  -\frac{1}{2T}\,\tau_0^{\langle\mu\nu\rangle}\,\delta h_{\mu\nu}
  -\frac{1}{T}\,S^\mu_\nu\,\delta\varepsilon_\mu^\nu\,.
\label{Gibbs-Duhem}
\end{align}

We define the {\em order} of a local quantity 
as the order of derivatives necessary to be taken 
in order to construct the quantity from local thermodynamic variables 
$p_\mu$, $n$, $g_{\mu\nu}$, and $\varepsilon_{\mu\nu}$\,:
\begin{align}
\begin{array}{ l | c }
  & \mbox{order} \\
 \hline
 p_\mu\,,~n\,,~g_{\mu\nu}\,,~\varepsilon_{\mu\nu}\,,~
  e\,,~u^\mu\,,~T\,,~\mu\,,~
  \tau_0^{\mu\nu}\,,~s ~& 0 \\
 \sigma_{\mu\nu}\,,~\omega_{\mu\nu}\,,\ldots & 1 \\
 \cR_{\mu\nu\lambda\sigma}\,,~\cR_{\mu\nu}\,,~
 \cR\,,~C_{\mu\nu\lambda\sigma}\,,\ldots & 2 \\
 ~~~\vdots & \vdots 
\end{array}
\end{align}
Accordingly, the energy-momentum tensor of a viscoelastic fluid 
has the following expansion in the Landau-Lifshitz frame: 
\begin{align}
 T^{\mu\nu} 
 &=e\,u^\mu u^\nu + \tau^{\mu\nu} \nn\\
 &=e\,u^\mu u^\nu + \tau_0^{\mu\nu} + \tau_1^{\mu\nu} + \tau_2^{\mu\nu} +\cdots
 \qquad\bigl(\tau_{i}^{\mu\nu} u_\nu=0\,;~i=0,1,2,\cdots\bigr)\,,
\end{align}
where the subscript $0,\,1,\,2,\ldots$ of a term stands for its order.

\section{Conformal viscoelastic fluids with large strain}
\label{conformal_large_strain}

\subsection{Definition}
\label{definition}

We say that a viscoelastic fluid is conformal 
if (1) its energy-momentum tensor is traceless $(T^\mu_\mu=0)$  
and has weight $w=d+2$ 
and (2) its equations of motion are Weyl covariant. 
We assume that the strain $\varepsilon=(\varepsilon_{\mu\nu})$ has 
the same weight as that of the metric $g_{\mu\nu}$ (i.e., $w=-2$). 
We further assume that the trace of $T^{\mu\nu}$ vanishes at each order, 
obtaining the equalities 
\begin{align}
 e=\tr \tau_0=(d-1)\,P\,,\qquad \tr \tau_i=0\quad(i=1,2,\ldots)\,. 
\end{align}
Thus, the conformal energy-momentum tensor is generically written 
in the following form:
\begin{align}
 T^{\mu\nu} &= e\, u^\mu u^\nu + \frac{e}{d-1}\,h^{\mu\nu}
  + \tau^{\langle\mu\nu\rangle} \nn\\
  &= e\, u^\mu u^\nu + \frac{e}{d-1}\,h^{\mu\nu} + \tau_0^{\langle\mu\nu\rangle}
  + \tau_1^{\langle\mu\nu\rangle} + \tau_2^{\langle\mu\nu\rangle} + \cdots\,.
\end{align}
To avoid discussions (and expressions) from being unnecessarily messy, 
we will set $\mu=0$ in what follows.%
\footnote{
For a conformal fluid (in the absence of strain), 
one can show from the Gibbs-Duhem relation \eqref{Gibbs-Duhem} 
that $e$ and $n$ are written in the form  
$e=\bar{e}\,T^k \mu^{d-k}$ 
and $n=\bigl((d-k)/(d-1)\bigr)\,\bar{e}\,T^k \mu^{d-k-1}$ 
with some constants $\bar{e}$ and $k$\,. 
} 

For a system of infinite size with free boundary condition, 
a system which has neither energy-momentum transfers 
nor strains at spatial infinity, 
a global equilibrium is characterized by the conditions that 
(i) $\partial \tilde{s}/\partial \tilde{p}_\mu = -u^\mu/T$ 
be spatially constant 
and 
(ii) $\varepsilon_{\mu\nu}=0$ \cite{fs1}. 
The Weyl-covariant expression for the condition (i) is given by 
\begin{align}
 0=h_{\mu}^{~\alpha}\,\cD_\alpha\Bigl(\frac{u^\nu}{T}\Bigr)
  =\frac{1}{T}\,\cD_\mu u^\nu
  + \Bigl[ h_{\mu}^{~\alpha}\,\cD_\alpha\Bigl(\frac{1}{T}\Bigr)\Bigr]\,u^\nu
  \qquad\mbox{(in global equilibrium)}\,,
\end{align}
or 
\begin{align}
 \left\{\begin{array}{ll} 
 \mbox{(i-a)} & \cD_\mu u_\nu = \sigma_{\mu\nu}+\omega_{\mu\nu} = 0 
\\
 \mbox{(i-b)} & h_{\mu}^{~\alpha}\,\cD_\alpha T
  ~\bigl(=h_\mu^{~\alpha}\,(\partial_\alpha T+a_\alpha\,T)\bigr)
  =0
  \end{array}\right.\quad\mbox{(in global equilibrium)}\,.
\end{align}
We know from (i-b) 
that the nonvanishing of spatial covariant derivatives of $T$ 
indicates departure of the system from global equilibrium,%
\footnote{
The acceleration $a_\mu=u^\nu\nabla_\nu u_\mu
=u^\nu\,(\partial_\nu u_\mu-\partial_\mu u_\nu)$ 
can be written as $a_\mu=N^{-1}\,h_\mu^{~\nu}\,\partial_\nu N$ 
for the ADM parametrization \eqref{ADM1} and \eqref{ADM2}. 
Thus, the condition (i-b) can be rewritten as 
$h_{\mu}^{~\alpha}\,\partial_\alpha (N\,T)=0$\,, 
which is equivalent to $\partial_i(N\,T)=0$\,. 
Note that $T_0\equiv N\,T$ is the temperature 
conjugate to the energy density measured with $x^0$ 
(not to the proper energy density measured with the local proper time) 
which includes the gravitational potential (Tolman's law), 
and that it is $T_0$ which becomes spatially constant in global equilibrium 
when the gravitational field exists \cite{fs1,Israel:1976tn,LL_statphys}.  
} 
so that it should be more natural to express 
thermodynamic quantities with the pair $(T,\,h,\,\varepsilon)$ 
than with $(e,\,h,\,\varepsilon)$ 
when making the derivative expansion around a global equilibrium state. 
The temperature $T$ is then the only dimensionful quantity 
for a conformal viscoelastic fluid, 
and we can express various thermodynamic quantities as follows 
[$\sigma=(\sigma_{\alpha\beta}),\,\omega=(\omega_{\alpha\beta})$]:
\begin{align}
 s&=T^{d-1}\,\bar{s}(h,\varepsilon)\,,\qquad
  e=T^d\,\bar{e}(h,\varepsilon)\,, \qquad
  S^\mu_\nu =T^d\,\bar{S}^\mu_\nu (h,\varepsilon)\,, \\
 \tau_0^{\mu\nu}&= T^d\,\bar\tau_0^{\mu\nu}(h,\varepsilon)\,,\qquad
  \tau_1^{\mu\nu}= 
  T^d\,\bar\tau_1^{\mu\nu}(h,\varepsilon,\,\sigma/T,\,\omega/T)\,,\quad\ldots
\label{scaling1}
\end{align}
From the fundamental relation \eqref{fundamental3} and 
the Euler relation \eqref{Euler} 
(with $\mu=0$ and $P=e/(d-1)$), 
we obtain the following equations:
\begin{align}
 \bar{s}&=\frac{d}{d-1}\,\bar{e}\,,\qquad
  \delta\bar{s}
  =\frac{d}{d-1}\,\delta\bar{e}
  =\frac{d}{2}\,\bar\tau_0^{\langle\mu\nu\rangle}\,\delta h_{\mu\nu}
  +d\,\bar{S}^\mu_\nu\,\delta\varepsilon_\mu^\nu\,.
\label{scaling2}
\end{align}

\subsection{Equations of motion}
\label{EOM}

The equations of motion comprise (1) the conservation law of $T^{\mu\nu}$ 
and (2) the rheology equations \cite{Azeyanagi:2009zd,Azeyanagi:2010ab,fs1,fs2}. 
The former, $\nabla_\mu T^{\mu\nu}=0$, 
determines the evolution of the energy-momentum $p^\mu=e\,u^\mu$ 
(or equivalently, that of $T$ and $u^\mu$) 
and describes the motion of material particles of a given material, 
while the latter determine the evolution of the strain 
$\varepsilon=(\varepsilon_{\mu\nu})$ 
and describe the plastic deformation of the material.

\noindent
\underline{(1) conservation law}

Since $T^{\mu\nu}$ is symmetric traceless and has weight $w=d+2$, 
the conservation law can be written in a Weyl-covariant form 
by using the equality \eqref{divergence_tensor} \cite{Loganayagam:2008is}:%
\footnote{
See \cite{fs1} for the derivation of the conservation law $\nabla_\mu T^{\mu\nu}=0$ 
on the basis of Onsager's linear regression theory. 
} 
\begin{align}
 \cD_\mu T^{\mu\nu} = \nabla_\mu T^{\mu\nu} =0 \,.
\end{align}

The longitudinal component can be further rewritten as follows: 
\begin{align}
 0&=-\,u_\nu\,\cD_\mu T^{\mu\nu}
  = -\cD_\mu(T^{\mu\nu} u_\nu) + T^{\mu\nu}\,\cD_\mu u_\nu \nn\\
 &= \cD_\mu (e u^\mu) + \Bigl[e\,u^\mu u^\nu + \frac{e}{d-1}\,h^{\mu\nu}
  + \tau^{\langle\mu\nu\rangle}\Bigr]\,\sigma_{\mu\nu}\nn\\
 &= \cD_u e + \tau^{\langle\mu\nu\rangle}\,\sigma_{\mu\nu}\,.
\label{cons_long}
\end{align}
This determines the time evolution of $e$\,. 
Setting $\delta=\cD_u$ in the Gibbs-Duhem relation \eqref{Gibbs-Duhem} 
[with $\mu=0$ and $P=e/(d-1)$], 
and using the identity $\cD_u g_{\mu\nu}=\cD_u h_{\mu\nu}=0$\,, 
we can also write down the equation determining the time evolution of $T$\,: 
\begin{align}
 \frac{\cD_u T}{T}&=\frac{1}{d}\,\frac{\cD_u e}{e}
  -\frac{d-1}{d}\,\frac{S^{\mu\nu}}{e}
  \,\cD_u\varepsilon_{\mu\nu}
\nn\\
 &=-\,\frac{1}{d\,\bar{e}}\,\bigl[
  \bar\tau^{\langle\mu\nu\rangle}\,\sigma_{\mu\nu}
  +(d-1)\,\bar{S}^{\mu\nu}\,\cD_u\varepsilon_{\mu\nu}\bigr]\,.
\label{zeroth}
\end{align}

The transverse component of the conservation law 
can be interpreted in two ways. 
When it is written with the standard covariant derivative $\nabla_\mu$\,, 
the equation $h^{\mu}_{~\nu}\cD_\alpha T^{\alpha\nu}=0$ gives the equation
\begin{align}
 a^\mu = u^\nu \nabla_\nu u^\mu 
  =-\,\frac{1}{e}\,h^{\mu}_{~\nu}\,\nabla_\alpha \tau^{\alpha\nu}\,,
\label{cons_trans}
\end{align}
which determines the time evolution of $u^\mu$ 
and describes how a material particle moves 
under the influence of the stress $\tau^{\mu\nu}$\,. 
The second interpretation is obtained 
when the equation is written in a manifestly Weyl-covariant form: 
\begin{align}
 0&=h^{\mu}_{~\nu}\,\cD_\alpha T^{\alpha\nu}
  =h^{\mu}_{~\nu}\,\cD_\alpha \tau^{\alpha\nu}
  =h^{\mu}_{~\nu}\,\cD_\alpha\bigl(T^d\, \bar\tau^{\alpha\nu})\,,
\end{align}
or
\begin{align}
 0&=h^{\mu}_{~\nu}\,\cD_\alpha \bar\tau^{\alpha\nu}
  + d\,\bar\tau^{\mu\nu}\,
  \Bigl(h_{\nu}^{~\alpha}\,\frac{\cD_\alpha T}{T}\Bigr) \,,
\label{DTspatial}
\end{align}
from which the spatial derivatives of the temperatures, 
$h_{\mu}^{~\alpha}\,\cD_\alpha T/T$\,, 
can be expressed in terms of other local variables.

Equations \eqref{zeroth} and \eqref{DTspatial} 
and the identity $\cD_u h_{\mu\nu}=0$ 
thus allow us to concentrate our consideration 
only on the time evolution of the strain.

\noindent
\underline{(2) rheology equations}

The rheology equations, on the other hand, 
have the following form in the derivative expansion:
\begin{align}
 \cD_u\varepsilon_{\mu\nu}
  =[\cD_u\varepsilon_{\mu\nu}]_0+[\cD_u\varepsilon_{\mu\nu}]_1
  +[\cD_u\varepsilon_{\mu\nu}]_2+\cdots\,.
\end{align}
Their explicit form will be given in section \ref{transient} 
assuming that the strain is small 
[see Eqs.~\eqref{rheology_eq1} and \eqref{rheology_eq2}].

\subsection{Entropy production and the second law of thermodynamics}
\label{entropy_second}

The entropy current $s^\mu$ consists of the convective part $s\,u^\mu$ 
and the dissipative part $s_{\rm (d)}^\mu$\,,
\begin{align}
 s^\mu = s\,u^\mu + s_{\rm (d)}^\mu\,,
\end{align}
from which the total entropy density is defined as 
\begin{align}
 s_{\rm tot}\equiv -u_\mu\,s^\mu=s+\Delta s
  \quad \bigl(\Delta s\equiv -u_\mu\,s_{\rm (d)}^\mu\bigr). 
\label{s_tot}
\end{align}
Under the hypothesis of local thermodynamic equilibrium 
(assumed throughout the present paper), 
$\Delta s$ is a local function of thermodynamic variables 
(not depending on their spatial derivatives) 
and can be absorbed into the original entropy density $s$\,. 
We, however, leave the possibility of the existence of $\Delta s$  
in order to make easier the comparison of the results to be obtained 
with those obtained in other references based on extended thermodynamics.

The entropy production rate is then given by 
\begin{align}
 \nabla_\mu s^\mu=\cD_\mu s^\mu=\cD_u s + \cD_\mu s_{\rm (d)}^\mu\,.
\label{entropy_production}
\end{align}
The first term of \eqref{entropy_production} can be expressed explicitly 
by using Eqs.~\eqref{fundamental3} and \eqref{cons_long} as
\begin{align}
 \cD_u s&= \frac{1}{T}\,\cD_u e+\frac{1}{T}\,S^\mu_\nu\,\cD_u \varepsilon_\mu^\nu
  =-\frac{1}{T}\,\tau^{\langle\mu\nu\rangle}\,\sigma_{\mu\nu}
  +\frac{1}{T}\,S^{\mu\nu}\,\cD_u\varepsilon_{\mu\nu}
\nn\\
 &= T^{d-1}\,\bigl(-\bar\tau^{\langle\mu\nu\rangle}\,\sigma_{\mu\nu}
  +\bar{S}^{\mu\nu}\,\cD_u\varepsilon_{\mu\nu}\bigr)\,,
\label{entropy_evol}
\end{align}
where we again have used the identity $\cD_u g_{\mu\nu}=\cD_u h_{\mu\nu}=0$\,.
The second term of \eqref{entropy_production} 
can be calculated once $s_{\rm (d)}^\mu$ is given explicitly, 
whose generic form in the derivative expansion is:
\begin{align}
 s_{\rm (d)}^\mu&=T^{d-1}\,( \bar{m}\,u^\mu + \bar{n}^\mu )
  \quad (\bar{n}^\mu u_\mu=0)\nn\\
 &=T^{d-1}\, \bigl[\, ( \bar{m}_0+\bar{m}_1+\bar{m}_2+\cdots)\,u^\mu
  + \bar{n}_0^\mu+\bar{n}_1^\mu+\bar{n}_2^\mu+\cdots\ \bigr]\,.
\label{entropy_diss}
\end{align}
Here, $\bar{n}_0^\mu=0$ 
since a dimensionless transverse vector cannot be constructed 
only from $h_{\mu\nu}$\,, $\varepsilon_{\mu\nu}$ and $u^\mu$\,. 
We will set $\bar{m}_0=0$ in the following discussions 
since $\bar{m}_0$ can be absorbed into $\bar{s}=s/T^{d-1}$\,. 
We leave the possibility of the existence of $\bar{m}_i$ ($i=1,2,\ldots$) 
for the reason stated below Eq.~\eqref{s_tot}.

\section{Conformal viscoelastic fluids at transient time scales}
\label{transient}

\subsection{Small strain expansion}
\label{small_strain}

We denote the shear (i.e.\ traceless) component of the strain tensor 
$\varepsilon=(\varepsilon_{\mu\nu})$ 
by $\varepsilon_S=(\varepsilon^S_{\mu\nu})
\equiv (\varepsilon_{\langle\mu\nu\rangle})$\,, 
and decompose the strain as 
\begin{align}
 \varepsilon_{\mu\nu}=\varepsilon^S_{\mu\nu}+\frac{1}{d-1}\,(\tr\varepsilon)\,
  h_{\mu\nu}\,.
\end{align}
The rheology equations then take the following form in the derivative expansion:%
\footnote{
An explicit parametrization is given 
in Eqs.~\eqref{rheology_eq1} and \eqref{rheology_eq2}.
}
\begin{align}
 \cD_u\varepsilon^S_{\mu\nu}
  &=\mbox{const.}\,T\,\varepsilon^S_{\mu\nu}
  +\mbox{const.}\,\sigma_{\mu\nu}+\cdots\,,
\label{rheology1}
\\
 \cD_u\tr\varepsilon
  &=\mbox{const.}\,T\,\tr\varepsilon+\mbox{const.}\,T\,\tr(\varepsilon_S^2)
  +\mbox{const.}\,\tr(\varepsilon^S\sigma)
  +\mbox{const.}\,T^{-1}\,\tr(\sigma^2)+\cdots\,.
\label{rheology2}
\end{align}
As time elapses, 
the strain gets relaxed and the rate of change becomes small. 
If the time scale of observation is sufficiently long, 
then the left hand sides of Eqs.~\eqref{rheology1} 
and \eqref{rheology2} become negligible compared to the right hand sides, 
and we obtain%
\footnote{
Note that, unlike the shear part \eqref{shear_part}, 
the trace part, $\tr\varepsilon$\,, 
cannot contain terms proportional to the gradient of velocity, 
$\vartheta=\nabla_\mu u^\mu$\,, in the derivative expansion, 
since $\vartheta$ is not a conformal scalar. 
This is consistent with the absence of bulk viscosity 
in conformal fluids.
} 
\begin{align}
 \varepsilon^S_{\mu\nu}&\,\rightarrow\,
  \mbox{const.}\,T^{-1}\,\sigma_{\mu\nu}+\cdots\,,\label{shear_part}\\
 \tr\varepsilon&\,\rightarrow\,
  \mbox{const.}\,\tr(\varepsilon_S^2)
  +\mbox{const.}\,T^{-1}\,\tr(\varepsilon_S\,\sigma)
  +\mbox{const.}\,T^{-2}\,\tr(\sigma^2) +\cdots \nn\\
  &\,\simeq\,\, \mbox{const.}\,T^{-2}\,\tr(\sigma^2)+\cdots\,.
\end{align}
Thus, in order to investigate a conformal viscoelastic fluid 
at the transient time scales 
where $\cD_u\varepsilon_{\mu\nu}$ is small but still not negligible, 
it is convenient to expand the equations of motion 
based on the following {\em semi-long time (SLT) order}\,:%
\begin{align}
\begin{array}{l|c}
 & \mbox{SLT-order} \\
 \hline
 e,\,T,\,\ldots & 0 \\
 \sigma_{\mu\nu},\,\omega_{\mu\nu},\,\varepsilon^S_{\mu\nu},\,\ldots & 1 \\
 \tr\varepsilon,\,\cR_{\langle\mu\nu\rangle},\,\ldots & 2 \\
 ~~~\vdots & \vdots
\end{array} 
\end{align}
For the rest of this section,
we investigate a conformal viscoelastic fluid 
at the transient time scales 
to second SLT-order in the constitutive equations. 

We first expand the energy-momentum tensor
\begin{align}
 T^{\mu\nu}
 &=T^d\,\Bigl[
  \bar{e}\,u^\mu u^\nu+\frac{\bar{e}}{d-1}\,h^{\mu\nu}
  + \bar\tau^{\langle\mu\nu\rangle} \Bigr]
\label{constitutive_eq1}
\end{align}
to second SLT-order as%
\footnote{
As can be seen from Eqs.~\eqref{uCu_uRu} and \eqref{dsigma_uCu}, 
only two are independent among the symmetric, transverse, traceless tensors 
of weight two; 
$\cD_u\sigma_{\mu\nu}$\,, 
$u^\alpha\cR_{\alpha\langle\mu\nu\rangle\beta}u^\beta$\,, 
$\cR_{\langle\mu\nu\rangle}$\,, 
$u^\alpha C_{\alpha\mu\nu\beta} u^\beta$ \cite{Loganayagam:2008is}. 
We will take $\cR_{\langle\mu\nu\rangle}$ 
and $u^\alpha C_{\alpha\mu\nu\beta} u^\beta$ 
as such two in the following discussions. 
} 
\begin{align}
 \bar\tau^{\langle\mu\nu\rangle}
  = &\,\, a_1\, T^{-1}\, \sigma^{\mu\nu}
    +\,a_2\, T^{-2}\, (\sigma^2)^{\langle\mu\nu\rangle}
    +a_3\, T^{-2}\, (\sigma\,\omega)^{\langle\mu\nu\rangle} 
    +a_4\, T^{-2}\, (\omega^2)^{\langle\mu\nu\rangle} \nn\\
 & +\,a_5\, T^{-2}\, u_\alpha \, C^{\alpha\langle\mu\nu\rangle\beta} \, u_\beta 
    +a_6\, T^{-2}\, \cR^{\langle\mu\nu\rangle}  \nn\\
 & +\,b_1\, \varepsilon_S^{\mu\nu} 
    +b_2\, \, (\varepsilon_S^2)^{\langle\mu\nu\rangle} 
    +b_3\, T^{-1}\, (\varepsilon_S\,\sigma)^{\langle\mu\nu\rangle} 
    +b_4\, T^{-1}\, (\varepsilon_S\,\omega)^{\langle\mu\nu\rangle}
   + \cdots \,.
\label{constitutive_eq2}
\end{align}
We also expand the entropic force caused by the strain, 
$S^{\mu\nu}=T^d\,\bar{S}^{\mu\nu}$\,, as
\begin{align}
 \bar{S}^{\mu\nu}
  =\bar{S}^{\langle\mu\nu\rangle}+\frac{1}{d-1}\,(\tr \bar{S})\,h^{\mu\nu}
  =\ell_1\,\varepsilon_S^{\mu\nu}
  + \ell_2\,(\varepsilon_S^2)^{\langle\mu\nu\rangle}+\cdots
  + \bigl(\ell_3\,\tr\varepsilon+\cdots\bigr)\,h^{\mu\nu}\,,
\end{align}
and the rheology equations as%
\footnote{
The equations generalize those given in \cite{Azeyanagi:2009zd,Azeyanagi:2010ab,fs2} 
(where only $c_1$ and $d_1$ are nonvanishing).
}
\begin{align}
 \cD_u\varepsilon_S^{\mu\nu}
 &= c_1\, \sigma^{\mu\nu} 
  +c_2\, T^{-1}\, (\sigma^2)^{\langle\mu\nu\rangle}
  +c_3\, T^{-1}\, (\sigma\,\omega)^{\langle\mu\nu\rangle}
  +c_4\, T^{-1}\, (\omega^2)^{\langle\mu\nu\rangle} \nn\\
 &~~~+\,c_5\, T^{-1}\, u_\alpha\, C^{\alpha\langle\mu\nu\rangle\beta} \,u_\beta 
  +c_6\, T^{-1}\, \cR^{\langle\mu\nu\rangle}  \nn\\
 &~~~+\,d_1\, T \, \varepsilon_S^{\mu\nu} 
  +d_2\, T \, (\varepsilon_S^2)^{\langle\mu\nu\rangle} 
  +d_3\, (\varepsilon_S\,\sigma)^{\langle\mu\nu\rangle} 
  +d_4\, (\varepsilon_S\,\omega)^{\langle\mu\nu\rangle}+\cdots \,,
\label{rheology_eq1}
\\
 \cD_u\tr\varepsilon&= f_0\,T\tr\varepsilon
  +f_1\,T\tr(\varepsilon_S^2)
  +f_2\,\tr(\varepsilon_S\, \sigma)
  +f_3\, T^{-1}\, \tr(\sigma^2)
  +f_4\, T^{-1}\, \tr(\omega^2) \nn\\
 &~~~+f_5\,T \tr(\varepsilon_S^3)
  +f_6\, \tr(\varepsilon_S^2 \sigma)
  +f_7\,T^{-1}\, \tr(\varepsilon_S\sigma^2)\nn\\
 &~~~+f_8\,T^{-1}\, \tr(\varepsilon_S\,\sigma\, \omega)
  +f_9\,T^{-1}\, \tr(\varepsilon_S\,\omega^2)
  +f_{10}\,T^{-2}\, \tr(\sigma^3)
  +f_{11}\,T^{-2}\, \tr(\sigma\,\omega^2) \nn\\
 &~~~+f_{12}\,T^{-1}\, \varepsilon_S^{\mu\nu}\, u^\alpha \, 
  C_{\alpha\langle\mu\nu\rangle\beta} \, u^\beta
  +f_{13}\,T^{-2}\, \sigma^{\mu\nu}\, 
  u^\alpha \, C_{\alpha\langle\mu\nu\rangle\beta} \, u^\beta\nn\\
 &~~~+f_{14}\,T^{-1}\, 
      \varepsilon_S^{\mu\nu}\, \cR_{\langle\mu\nu\rangle}
  +f_{15}\,T^{-2}\, \sigma^{\mu\nu}\, \cR_{\langle\mu\nu\rangle}\nn\\
 &~~~+f_{16}\,T^{-1}\,\cD_\mu\cD_\nu \varepsilon_S^{\mu\nu}
  +f_{17}\,T^{-2}\,\cD_\mu\cD_\nu \sigma^{\mu\nu} +\cdots\,.
\label{rheology_eq2}
\end{align}
Then, by using Eq.~\eqref{entropy_evol}, 
$\cD_u s$ can be expressed explicitly as follows:%
\footnote{
Since $(\varepsilon_S)^{\rm T}=\varepsilon_S
\,,\,\sigma^{\rm T}=\sigma\,,\,\omega^{\rm T}=-\omega$\,,
we have $\tr(\varepsilon_S\sigma\omega)
=(1/2)\,\tr\bigl((\varepsilon_S\sigma+\sigma\varepsilon_S)\omega\bigr)
=(1/2)\,\tr\bigl(\varepsilon_S(\sigma\omega-\omega\sigma)\bigr)$\,.
}
\begin{align}
 \cD_u s 
 = &\,\, T^d\,\bigl[\,
  \ell_1\, d_1 \, \tr(\varepsilon_S^2)
  + (-b_1+\ell_1\,c_1)\, T^{-1}\, \tr(\varepsilon_S\, \sigma)
  - a_1 \, T^{-2}\, \tr(\sigma^2) 
\nn\\
 &+ (\ell_2\,d_1+\ell_1\,d_2) \, \tr(\varepsilon_S^3) 
  + (-b_2+\ell_2\,c_1+\ell_1\,d_3)\, T^{-1}\, \tr(\varepsilon_S^2\, \sigma) 
\nn\\
 &+ (-b_3+\ell_1\,c_2)\, T^{-2}\, \tr(\varepsilon_S\,\sigma^2) 
  + (b_4+\ell_1\,c_3)\, T^{-2}\, \tr(\varepsilon_S\,\sigma\, \omega)
  + \ell_1\,c_4 \, T^{-2}\, \tr(\varepsilon_S\,\omega^2) 
\nn\\
 &-a_2 \, T^{-3}\, \tr(\sigma^3)
  - a_4 \, T^{-3}\, \tr(\sigma\,\omega^2) \
  + \ell_1\,c_5 \, T^{-2}\, \varepsilon_S^{\mu\nu}\, 
  u^\alpha \, C_{\alpha\langle\mu\nu\rangle\beta} \, u^\beta
\nn\\
 & - a_5\, T^{-3}\, \sigma^{\mu\nu}\, u^\alpha \, 
  C_{\alpha\langle\mu\nu\rangle\beta} \, u^\beta 
  + \ell_1\,c_6 \, T^{-2}\, 
  \varepsilon_S^{\mu\nu}\, \cR_{\langle\mu\nu\rangle}
  - a_6 \, T^{-3}\, \sigma^{\mu\nu}\, \cR_{\langle\mu\nu\rangle} \,\bigr]\,.
\end{align}

As for the dissipative part $s_{\rm (d)}^\mu$ in the entropy current 
$s^\mu=s\,u^\mu+s_{\rm (d)}^\mu$\,, 
we need to expand it to second SLT-order:%
\footnote{
We have neglected terms proportional 
to $T^{d-1}\,u^\mu \tr\varepsilon$ 
or $T^{d-1}\,u^\mu \tr(\varepsilon_S^2)$ 
since they can be absorbed into $s$ 
[see a comment below Eq.~\eqref{entropy_diss}]. 
We also have neglected terms proportional to $\cF^{\mu\nu} \,u_\nu$ 
since it is at least of third-order derivative 
[see Eq.~\eqref{third-order}].
}
\begin{align}
 s^\mu_{\rm (d)} 
  = & \,\, T^{d-1}\,\bigl\{ \,\bigl[\,
  A_1    \,T^{-1}\,\tr(\varepsilon_S\,\sigma)
  + (A_2/2) \,T^{-2}\,\tr(\sigma^2)
  + (A_3/2) \,T^{-2}\,\tr(\omega^2) 
\nn\\
 &
  + (A_4/2)\,T^{-1}\,\tr(\varepsilon_S^2\,\sigma) 
  + A_5 \,T^{-2}\,\tr(\varepsilon_S\,\sigma^2) 
  + A_6 \,T^{-2}\,\tr(\varepsilon_S\,\sigma\,\omega) 
\nn\\
 & + A_7 \,T^{-2}\,\tr(\varepsilon_S\,\omega^2) 
   + A_8 \,T^{-2}\,\cR +\cdots\,\bigr]\,u^\mu \nn\\
 & + A_9 \,T^{-1}\, \cD_\nu\varepsilon_S^{\mu\nu} 
   + A_{10} \,T^{-2}\, \cD_\nu\sigma^{\mu\nu} 
   + A_{11} \,T^{-2}\, \cD_\nu\omega^{\mu\nu} 
   + A_{12} \,T^{-2}\, u_\nu \cR^{\nu\rho} h_\rho^{~\mu}+\cdots
   \,\bigr\}\,. 
\end{align}
As mentioned before, 
the scalar in front of $u^\mu$ 
represents the correction $\Delta s$ to the original entropy density $s$\,;  
$\Delta s=s_{\rm tot}-s$.
Local thermodynamic equilibrium is inevitably broken 
when any of these coefficients $A_i$ $(i=1,\ldots,8)$ do not vanish. 
Although we eventually set $A_i=0$ $(i=1,\ldots,8)$ later, 
we leave them for a while for comparison with other references.

The entropy production rate can then be written in the following form 
[note that $\cD_\mu s^\mu = \nabla_\mu s^\mu$ 
due to Eq.\ \eqref{divergence_vector}]: 
\begin{align}
 &\cD_\mu s^\mu = \cD_\mu \bigl(s\, u^\mu + s^\mu_{\rm (d)}\bigr)
  =\cD_u s + \cD_\mu s_{\rm (d)}^\mu \nn\\
 &= T^d\,\Bigl\{\, 
  \ell_1\,d_1\, \tr(\varepsilon_S^2) 
  +\,(-b_1+\ell_1\,c_1+A_1\,d_1)\, 
  T^{-1}\, \tr(\varepsilon_S\, \sigma) 
  +(-a_1+A_1\,c_1)\, T^{-2}\, \tr(\sigma^2)
\nn\\
 & ~~~
  +(\ell_2\,d_1+\ell_1\,d_2)\,\tr(\varepsilon_S^3) 
  +(-b_2 + \ell_2\,c_1+A_1\,d_2+A_4\,d_1+\ell_1\,d_3)\,
  T^{-1}\, \tr(\varepsilon_S^2\,\sigma) 
\nn\\
 & ~~~
  +(-b_3+A_4\,c_1+\ell_1\,c_2+A_5\,d_1+A_1\,d_3-A_1)\, T^{-2}\, 
  \tr(\varepsilon_S\,\sigma^2) \nn\\
 & ~~~
  +(b_4+\ell_1\,c_3+A_6\,d_1-A_1\,d_4)\,
     T^{-2}\, \tr(\varepsilon_S\,\sigma\, \omega)  
  +(\ell_1\,c_4+A_7\,d_1-A_1)\, 
  T^{-2}\, \tr(\varepsilon_S\,\omega^2) 
\nn\\
 & ~~~
  +(-a_2 + A_1\,c_2 +A_5\,c_1-A_2+2 A_{12})\, 
  T^{-3}\, \tr(\sigma^3) 
\nn\\
 & ~~~
  +(-a_4 + A_7\,c_1 + A_1\,c_4- A_2 - 2 A_3+6 A_{12})\, 
  T^{-3}\, \tr(\sigma\,\omega^2) 
\nn\\ 
 & ~~~
  +(\ell_1\,c_5 +A_1)\, 
  T^{-2}\, \varepsilon_S^{\mu\nu}\, u^\alpha \, 
  C_{\alpha\langle\mu\nu\rangle\beta} \, u^\beta 
  +(-a_5+A_1\,c_5+A_2-2 A_{12})\, T^{-3}\,
  \sigma^{\mu\nu}\, 
  u^\alpha \, C_{\alpha\langle\mu\nu\rangle\beta} \, u^\beta \nn\\
 & ~~~
  +\Bigl(\ell_1\,c_6+\frac{A_1}{d-2}\Bigr)\, 
  T^{-2}\,\varepsilon_S^{\mu\nu}\, \cR_{\langle\mu\nu\rangle} \nn\\
 & ~~~
  +\Bigl(-a_6+A_1\,c_6+\frac{A_2}{d-2}-2 A_8-\frac{2}{d-2}\,A_{12} \Bigr)\, 
  T^{-3}\, \sigma^{\mu\nu}\, \cR_{\langle\mu\nu\rangle} 
  + A_9\,T^{-2}\,
  \cD_\mu\cD_\nu \varepsilon_S^{\mu\nu} 
\nn\\
 & ~~~
  + (2 A_8 +A_{10} +A_{12})\,T^{-3}\,
  \cD_\mu\cD_\nu \sigma^{\mu\nu} 
  + \cdots
 \,\Bigr\} \,.
\end{align}
Here we have used Eq.~\eqref{D_uRR} 
and neglected terms proportional to 
$\cD^\mu(\cF_{\mu\nu}\,u^\nu)$ 
or 
$\cD_\mu\cD_\nu\omega^{\mu\nu}\allowbreak 
=-\bigl((d-3)/2\bigr)\cF_{\mu\nu}\,\omega^{\mu\nu}$ 
since they are at least forth-order derivatives   
[see Eqs.~\eqref{third-order} and \eqref{third-order2}]. 
We also have used the fact that $T$ can be treated 
as being covariantly constant to this order ($\cD_\mu T\simeq 0$) 
since both $\cD_u T$ and $h_{\mu}^{~\alpha}\,\cD_\alpha T$ are 
at least of second SLT-order 
[see Eqs.~\eqref{zeroth} and \eqref{DTspatial}].

Now, if we set 
\begin{align}
 A_9 =0 \,, \qquad
  2\,A_8+A_{10}+A_{12}=0 \,,  
\end{align}
the entropy production rate takes the following bilinear form 
in our approximation: 
\begin{align}
 \cD_\mu s^\mu 
 =  T^{d}\,\vec{V}^{\,{\rm T}}_{\mu\nu}\, \cM \,\vec{V}^{\mu\nu} 
\end{align}
with
\begin{align}
 \cM&=\begin{pmatrix}
  \ell_1\,d_1  & 
   \frac{\ell_1\,c_1-b_1+A_1\,d_1}{2} & * \cr
  \frac{\ell_1\,c_1-b_1+A_1\,d_1}{2} & 
    -a_1+A_1\,c_1 & * \cr
  * & * & \boldsymbol{*}_{8\times 8} 
 \end{pmatrix} \,, 
\qquad
  \vec{V}^{\mu\nu} 
  &= \left(\begin{smallmatrix}
  \varepsilon_S^{\mu\nu} \cr
  \sigma^{\mu\nu}/T \cr
  (\varepsilon_S^2)^{\langle\mu\nu\rangle} \cr
  (\varepsilon_S\,\sigma)^{\langle\mu\nu\rangle}/T \cr
  (\varepsilon_S\,\omega)^{\langle\mu\nu\rangle}/T \cr
  (\sigma^2)^{\langle\mu\nu\rangle}/T^2 \cr
  (\sigma \, \omega)^{\langle\mu\nu\rangle}/T^2 \cr
  (\omega^2)^{\langle\mu\nu\rangle}/T^2 \cr
  u_\alpha \, C^{\alpha\langle\mu\nu\rangle\beta} \, u_\beta/T^2 \cr
  \cR^{\langle\mu\nu\rangle}/T^2 
  \end{smallmatrix}\right) \,. 
\end{align}
The second law of thermodynamics is then certified 
if the coefficient matrix $\cM$ is positive semi-definite.

The conformal fluid mechanics of \cite{Loganayagam:2008is} 
is obtained by setting the parameters as follows:
\begin{align}
 a_1 &= \eta_1\,,\quad 
 a_2 = -\eta_2+\eta_4\,,\quad 
 a_3 = -2\,\eta_3\,,\quad 
 a_4 = -\eta_2+\eta_5\,,\nn\\
 a_5 &= \eta_2+\eta_6 \,, \quad 
 a_6 = \frac{\eta_2}{d-2}\,,\quad 
 b_i=c_i=d_i=f_i=0\quad (i=1,2,\cdots)\,,\nn\\
 A_2&=\eta_2+\frac{d-4}{d-2}\,\eta_6\,,\quad 
 A_3 =-\frac{1}{2}\,\Bigl(\eta_5 + \frac{d+2}{d-2}\,\eta_6\Bigr) \,,\quad 
 A_8 = \frac{\eta_6}{2(d-2)} \,,\quad
 A_{11} = \frac{\eta_5+3\eta_6}{2(d-3)} \,,\nn\\
 A_{12} &= -\frac{\eta_6}{d-2} \,, \qquad
 A_1 =A_4=A_5=A_6=A_7=A_9=A_{10}=0 \,.
\end{align}
A calculation based on the fluid/gravity correspondence 
shows that $A_2\neq0$\,, $A_3\neq0$\,, $A_8\neq0$ 
\cite{Baier:2007ix,Loganayagam:2008is,Bhattacharyya:2008xc},
and thus local thermodynamic equilibrium is broken for such conformal fluids. 
In the next subsection, 
we show that the breakdown of local thermodynamic equilibrium can be avoided 
if a conformal fluid is always defined 
as the long time limit of a conformal viscoelastic system.

\subsection{Long time limit and second-order fluid mechanics}
\label{long_time_limit}

We now consider a conformal higher-order viscoelastic system 
when the time scale of observation is much longer than the relaxation times 
of the strain:
\begin{align}
 \cD_u\varepsilon_S^{\mu\nu} \ll d_1\, T\,\varepsilon_S^{\mu\nu} \,, \qquad
 \cD_u\tr\varepsilon \ll f_0\, T\tr\varepsilon \,.
\end{align}
As in first-order viscoelastic systems (see, e.g., \cite{fs2}), 
our second-order viscoelastic system comes to behave 
as a second-order viscous fluid (which is conformal now). 
In fact, the rheology equations \eqref{rheology_eq1} and \eqref{rheology_eq2} 
gives the following equations in the long time limit:%
\footnote{
Note that the long time limit of shear strain, Eq.~\eqref{long_time1}, 
has the same form as the additional dynamical variable 
$\xi_{\mu\nu}$ in divergence-type conformal fluid mechanics 
(see Eq.~(98) of \cite{PeraltaRamos:2009kg}). 
} 
\begin{align}
 \varepsilon_S^{\mu\nu} 
 \,\to\, &-\frac{1}{d_1}\,\bigl[\,
  c_1\, T^{-1}\, \sigma^{\mu\nu}  
  +c_2\, T^{-2}\, (\sigma^2)^{\langle\mu\nu\rangle} 
  +c_3\, T^{-2}\, (\sigma\,\omega)^{\langle\mu\nu\rangle} \nn\\
 & +c_4\, T^{-2}\, (\omega^2)^{\langle\mu\nu\rangle} 
  +c_5\, T^{-2}\, u_\alpha \, C^{\alpha\langle\mu\nu\rangle\beta} 
  \, u_\beta 
  +c_6\, T^{-2}\, \cR^{\langle\mu\nu\rangle} \nn\\
 & +d_2\, (\varepsilon_S^2)^{\langle\mu\nu\rangle}  
  +d_3\, T^{-1}\,(\varepsilon_S\, \sigma)^{\langle\mu\nu\rangle} 
  +d_4\, T^{-1}\,(\varepsilon_S\, \omega)^{\langle\mu\nu\rangle} 
  + \cdots \bigr]\nn\\
 \simeq\,\, & -\frac{c_1}{d_1}\,T^{-1}\,\sigma^{\mu\nu}
  -\Bigl[\frac{c_2}{d_1}+\frac{d_2}{d_1}\Bigl(\frac{c_1}{d_1}\Bigr)^2
   -\frac{d_3 c_1}{d_1^2}\Bigr]\,T^{-2}\,(\sigma^2)^{\langle\mu\nu\rangle} 
  - \Bigl(\frac{c_3}{d_1}-\frac{d_4 c_1}{d_1^2}\Bigr)\,T^{-2}\,
  (\sigma\omega)^{\langle\mu\nu\rangle}\nn\\
 & -\frac{c_4}{d_1}\,T^{-2}\,(\omega^2)^{\langle\mu\nu\rangle} 
 - \frac{c_5}{d_1}\,T^{-2}\,
  u_\alpha \, C^{\alpha\langle\mu\nu\rangle\beta} \, u_\beta
  -\frac{c_6}{d_1}\,T^{-2}\,\cR^{\langle\mu\nu\rangle}+\cdots \,,
\label{long_time1}
\\
 \tr\varepsilon \,\to\, & -\frac{1}{f_0}\,\bigl[\,
  f_1\,\tr(\varepsilon_S^2)
  +f_2\,T^{-1}\,\tr(\varepsilon_S\sigma)
  +f_3\,T^{-2}\,\tr(\sigma^2)+f_4\,T^{-2}\,\tr(\omega^2)
  +\cdots \bigr]\nn\\
 \simeq\,\, & -\Bigl[\frac{f_1}{f_0}\,\Bigl(\frac{c_1}{d_1}\Bigr)^2
  -\frac{f_2}{f_0}\frac{c_1}{d_1}+\frac{f_3}{f_0}\,\Bigr]\,T^{-2}\,\tr(\sigma^2)
  -\frac{f_4}{f_0}\,T^{-2}\,\tr(\omega^2)+\cdots\,.
\label{long_time2}
\end{align}
Substituting this into the constitutive equations \eqref{constitutive_eq1} 
and \eqref{constitutive_eq2}, 
we obtain the energy-momentum tensor in the long time limit: 
\begin{align}
 T^{\mu\nu}_{\rm (long)}
 =&\,\, a_0\,  T^d\, (g^{\mu\nu}+d\, u^\mu u^\nu) 
  + \frac{a_1\,d_1-b_1\,c_1}{d_1}\, T^{d-1}\, \sigma^{\mu\nu} \nn\\
 & + \frac{a_2\,d_1^3 + b_1\,(c_1\,d_3\,d_1 - c_1^2\,d_2 - c_2\,d_1^2) 
  + b_2\,c_1^2\,d_1 - b_3\,c_1\,d_1^2}{d_1^3} 
  \, T^{d-2}\, (\sigma^2)^{\langle\mu\nu\rangle} \nn\\
 & +\frac{a_3\,d_1^2+b_1\,(c_1\,d_4-c_3\,d_1)-b_4\,c_1\,d_1}{d_1^2}\, 
  T^{d-1}\, (\sigma\, \omega)^{\langle\mu\nu\rangle}
   + \frac{a_4\,d_1-b_1\,c_4}{d_1} \, T^{d-2}\, 
  (\omega^2)^{\langle\mu\nu\rangle} \nn\\
 & + \frac{a_5\,d_1-b_1\,c_5}{d_1}\, T^{d-2}\, 
  u_\alpha \, C^{\alpha\langle\mu\nu\rangle\beta} \, u_\beta 
  + \frac{a_6\,d_1-b_1\,c_6}{d_1}\, T^{d-2}\, \cR^{\langle\mu\nu\rangle} 
  +\cdots \,.
\label{constitutive_eq_long}
\end{align}
Here, we have set $\bar{e}=(d-1)\,a_0$ ($a_0$: constant) 
since $\bar{e}$ becomes constant in the long time limit 
(Stefan-Boltzmann law). 
Equation \eqref{constitutive_eq_long} has 
the same form as the energy-momentum tensor 
of a generic conformal second-order fluid 
\cite{Loganayagam:2008is,Bhattacharyya:2008xc,Bhattacharyya:2008mz}.

Interestingly, even if we start from a viscoelastic system 
with manifest local thermodynamic equilibrium 
[i.e.\ a system with $A_i=0$ $(i=1,\ldots,8)$ so that $\Delta s =0$], 
all terms can appear in $T^{\mu\nu}_{\rm (long)}$ with any possible values, 
in a way consistent with the second law of thermodynamics. 
In fact, there are no constraints on parameters in the constitutive equations 
in order for the strain $\varepsilon_{\mu\nu}$ to be converted 
to spatial derivatives (such as $\sigma_{\mu\nu}$ and $\omega_{\mu\nu}$) 
in the long time limit 
[see Eqs.~\eqref{long_time1} and \eqref{long_time2}]. 
Furthermore, we can also understand the appearance of spatial derivatives 
in the entropy density of a conformal fluid 
as a result of the same conversion mechanism, 
which is now applied to the entropy density 
with manifest local thermodynamic equilibrium:  
\begin{align}
 s &= s(p_\mu,\, g_{\mu\nu},\,\varepsilon_{\mu\nu})
=s(p_\mu,\,g_{\mu\nu},\,0)
  + \mbox{const.}\,T^{d-1}\,\tr(\varepsilon_S^2)
  + \mbox{const.}\,T^{d-1}\,\tr\varepsilon + \cdots\,,
\end{align}
that transmutes in the long time limit 
into the one with spatial derivatives: 
\begin{align}
 s_{\rm (long)}&=
  s(p_\mu,\,g_{\mu\nu},\,0) + \mbox{const.}\,\,T^{d-3}\,\tr(\sigma^2)
  +\mbox{const.}\,\,T^{d-3}\,\tr(\omega^2)
  +\mbox{const.}\, T^{d-3}\,\cR +\cdots \,. 
\label{transmutation}
\end{align}
Thus, even though the hypothesis of local thermodynamic equilibrium 
holds at short time scales for our viscoelastic system, 
it is seemingly broken when the system is observed at long time scales 
and is treated as a viscous fluid.

\section{Conclusion and discussions}
\label{conclusion}

In this paper, 
we defined conformal higher-order viscoelastic fluid mechanics. 
We wrote down the equations of motion  
in such a way that the evolution is consistent 
with the second law of thermodynamics. 
We further showed that
any conformal second-order fluid
with arbitrary parameters in the constitutive equations
can be obtained
by taking the long time limit of a viscoelastic conformal fluid, 
without violating the hypothesis of local thermodynamic equilibrium.

On the other hand, 
if one trusts the fluid/gravity correspondence, 
the entropy current $s^\mu$\ of a conformal fluid  
can be computed in the gravity side. 
The result \cite{Loganayagam:2008is,Bhattacharyya:2008xc} shows that 
the total entropy density $s_{\rm tot}$ contains spatial derivative terms 
with nonvanishing coefficients, 
and thus we know that local thermodynamic equilibrium is violated 
even at short distance scales 
for such conformal fluids that have gravity duals.%
\footnote{
One should be careful about this statement.
In fact, the local entropy current is defined
in the fluid/gravity correspondence
by pulling-back the area form on the horizon 
to the boundary on which fluid mechanics is defined. 
However, there are ambiguities in the definition \cite{Bhattacharyya:2008xc} 
(e.g., the ambiguity in making the boundary-to-horizon map, 
and the ambiguity of whether the area form is constructed 
on the event horizon or on the apparent horizon). 
Thus, it might be possible to find a suitable definition 
of the local entropy such that local thermodynamic equilibrium is not violated. 
} 

As was argued in \cite{fs1},
even when local thermodynamic equilibrium is realized 
for a system with resolution $(\epsilon_{\rm t},\,\epsilon_{\rm s})$, 
spatial derivative terms are naturally induced in the total entropy density 
(as the {\em entropy functional} in the language of \cite{fs1} 
or as in Eq.~\eqref{transmutation})
if we observe the system at larger scales 
in both the temporal and the spatial directions. 
Since viscoelastic fluids allow a description  
with manifest local thermodynamic equilibrium, 
we expect that the fluid/gravity correspondence is 
an already coarse-grained correspondence 
between viscoelastic fluid mechanics and 
a more microscopic description of gravity. 
If this is the case, 
it then should give an important clue to finding fundamental degrees of freedom 
in quantum gravity 
to try to formulate such ``viscoelasticity/quantum gravity correspondence.'' 
A study along this line is now in progress 
and will be reported elsewhere \cite{fs4}. 

As another direction of future research, 
it would be interesting to apply viscoelastic fluid mechanics 
to the phenomenology of heavy-ion collision experiments. 
In fact, relativistic viscoelastic model 
gives a causal completion of relativistic fluid mechanics
(the latter being defined as the long time limit of the former) \cite{fs2}, 
and thus it is tempting to assume that 
there is a phase of viscoelasticity 
prior to the stage of viscous fluidity. 
Then, it is important to investigate 
how elasticity at short time scales 
affects the dynamics of states right after collisions. 
In particular, one should investigate 
whether elasticity drives the system 
to an ideal fluid more rapidly than in the standard second-order fluid mechanics, 
as has been observed in divergence-type fluid mechanics \cite{PeraltaRamos:2009kg}. 

\section*{Acknowledgments}

The authors thank Hikaru Kawai, Teiji Kunihiro 
and Makoto Natsuume for useful discussions. 
They also thank Jeronimo Peralta-Ramos 
to bring their attention to reference \cite{PeraltaRamos:2009kg}. 
This work was supported by the Grant-in-Aid for the Global COE program 
``The Next Generation of Physics, Spun from Universality and
Emergence" from the Ministry of Education, Culture, Sports, 
Science and Technology (MEXT) of Japan. 
This work was also supported by the Japan Society for the Promotion of Science 
(JSPS) (Grant No.\,21$\cdot$1105) and by MEXT (Grant No.\,23540304).

\appendix

\section{Weights of local thermodynamic variables}
\label{weights}
\label{weight_list}

The Weyl weight of a $(p,q)$ tensor 
$Q^{\mu_1\cdots\mu_p}_{\nu_1\cdots\nu_q}$ 
of dimension $\Delta$ is given by 
$w=\Delta+p-q$\,.
We list below the dimensions and the weights 
of various local thermodynamic quantities. 

\begin{align}
\begin{array}{ l||c|c|c|c }
  & \mbox{dimension $\Delta$} & \mbox{weight $w$} & \mbox{order}
   & \mbox{SLT-order}\\
 \hline \hline
 \cD_\mu & 1 & 0 & 1 & \mbox{N/A} \\
 \hline 
 u^\mu & 0 & 1 & 0 & 0\\
 \varepsilon^S_{\mu\nu}=\varepsilon_{\langle\mu\nu\rangle} & 0 & -2 & 0 & 1 \\
 \tr\varepsilon & 0 & 0 & 0 & 2\\
 g_{\mu\nu} & 0 & -2 & 0 & 0\\
 h_{\mu\nu}=g_{\mu\nu}+u_\mu u_\nu & 0 & -2 & 0 & 0\\
 \hline 
 \cR_{\mu\nu\lambda\sigma} & 2 & -2 & 2 \\
 C_{\mu\nu\lambda\sigma} & 2 & -2 & 2 \\
 \cR_{\mu\nu}= \cR_{\mu\alpha\nu}{}^{\alpha} & 2 & 0 & 2 \\
 \cF_{\mu\nu}=\partial_\mu \cA_\nu-\partial_\nu \cA_\mu & 2 & 0 & 2 \\
 \cR=\cR_{\alpha}^{~\alpha} & 2 & 2 & 2 \\
 \hline
 p_\mu & d & d-1 & 0 & 0 \\
 e=\sqrt{-g^{\mu\nu} p_\mu p_\nu}  & d & d & 0 & 0 \\
 s=s(e,\varepsilon_{\mu\nu}) & d-1 & d-1 & 0 & 0 \\
 T=(\partial s/\partial e)^{-1} & 1 & 1 & 0 & 0 \\
 s^\mu=s\,u^\mu + s_{\rm (d)}^\mu & d-1 & d & \mbox{N/A} & \mbox{N/A} \\
 T^{\mu\nu}=e\,u^\mu u^\nu + \tau^{\mu\nu} & d & d+2 
  & \mbox{N/A} & \mbox{N/A}
\end{array}
\end{align}

\section{Useful formulas}
\label{formulas}

In this appendix, we prove a few useful formulas 
which are used in the main text.

For the Weyl-covariantized Riemann tensor 
\eqref{Riemann_tensor_conformal}--\eqref{A_F}, 
the following equality holds:
\begin{align}
 u^\alpha\,u^\beta\,h_{\mu}^{~\rho}h_{\nu}^{~\sigma}\, 
  \cR_{\rho\alpha\sigma\beta} 
 &=u^\alpha\,u^\beta\,h_{\nu}^{~\sigma}\, \cR_{\mu\alpha\sigma\beta} 
  =u^\alpha \cR_{\mu\alpha\nu}{}^{\sigma} u_\sigma
  + u^\alpha u^\beta u^\sigma u_\nu \cR_{\mu\alpha\,(\sigma\beta)}
\nn\\
 &= u^\alpha \bigl(\cR_{\mu\alpha\nu}{}^{\sigma}
  -\cF_{\mu\alpha}\,\delta^\sigma_\nu\bigr)\, u_\sigma 
  =  u^\alpha\,[\cD_\mu, \cD_\alpha]\, u_\nu 
  = -\,\cD_\mu u^\alpha\,\cD_\alpha u_\nu - \cD_u\cD_\mu u_\nu
\nn\\
 &= -\,(\sigma^2)_{\mu\nu} 
     -(\sigma\,\omega)_{\mu\nu} 
     -(\omega\,\sigma)_{\mu\nu} 
     -(\omega^2)_{\mu\nu} 
     - \cD_u(\sigma_{\mu\nu}+\omega_{\mu\nu}) \,, 
\label{uuhR}
\end{align}
where $h_{\mu\nu}=g_{\mu\nu}+u_\mu u_\nu$ and $\cD_u= u^\mu\,\cD_\mu$\,. 
By decomposing Eq.~\eqref{uuhR} into the trace part, the symmetric traceless part, 
and the antisymmetric part, 
we obtain the following identities:%
\footnote{
We can show that Eqs.~\eqref{u_cR_u_trace}--\eqref{u_cR_u_anti} 
are equivalent to the well-known evolution equations of $\vartheta$, $\sigma_{\mu\nu}$, 
and $\omega_{\mu\nu}$ (such as the Raychaudhuri equation) 
by using the following equations:
\begin{align*}
 u^\alpha\,\cR_{\alpha\beta} \,u^\beta 
  &= u^\alpha\, R_{\alpha\beta} \,u^\beta 
     +\nabla_u\vartheta - \nabla_\mu a^\mu +\frac{\vartheta^2}{d-1} \,,
\\
 \cD_u \sigma_{\mu\nu} 
  &= h_{\mu}^{~\alpha} h_{\nu}^{~\beta} \cD_u \sigma_{\alpha\beta}
   = h_{\mu}^{~\alpha} h_{\nu}^{~\beta} \nabla_u \sigma_{\alpha\beta} 
      + \frac{\vartheta}{d-1}\, \sigma_{\mu\nu} \,,
\\
 \cD_u \omega_{\mu\nu} 
  &= h_{\mu}^{~\alpha} h_{\nu}^{~\beta} \cD_u \omega_{\alpha\beta}
   = h_{\mu}^{~\alpha} h_{\nu}^{~\beta} \nabla_u \omega_{\alpha\beta}
    + \frac{\vartheta}{d-1}\, \omega_{\mu\nu} \,, \\
 \frac{1}{2}\,h_{\mu}^{~\rho}\,h_{\nu}^{~\sigma}\,\cF_{\rho\sigma}  
 &= h_{\mu}^{~\alpha} h_{\nu}^{~\beta} \nabla_{[\alpha} a_{\beta]} 
   -\frac{\vartheta}{d-1}\,\omega_{\mu\nu} \,.
\end{align*}
}
\begin{align}
 u^\alpha\,\cR_{\alpha\beta} \,u^\beta
  &= -\,\tr(\sigma^2) - \tr(\omega^2) \,, 
\label{u_cR_u_trace}\\
 u^\alpha \, \cR_{\alpha\langle\mu\nu\rangle\beta}\, u^\beta 
  &= (\sigma^2)_{\langle\mu\nu\rangle} 
    +(\omega^2)_{\langle\mu\nu\rangle} + \cD_u \sigma_{\mu\nu} \,, 
 \label{u_cR_u_symm} \\
 \frac{1}{2}\,h_{\mu}^{~\rho}\,h_{\nu}^{~\sigma}\,\cF_{\rho\sigma} 
 &= (\sigma\,\omega+\omega\,\sigma)_{\mu\nu} + \cD_u \omega_{\mu\nu} \,,
 \label{u_cR_u_anti}
\end{align}
where we have used the relation 
$\cR_{\mu\alpha\,(\sigma\beta)}=\cF_{\mu\alpha}\, g_{\sigma\beta}$ 
and $u^\alpha \, \cR_{\alpha[\mu\nu]\beta}\, u^\beta = (1/2)\, \cF_{\mu\nu}$ 
[see Eqs.~\eqref{indices1} and \eqref{indices3}].

The Weyl tensor \eqref{Weyl_tensor},
\begin{align}
 C_{\mu\nu\lambda\sigma}&\equiv 
  R_{\mu\nu\lambda\sigma}
  +\frac{4}{d-2}\,\,\delta^\alpha_{[\mu}g_{\nu][\lambda}\delta^\beta_{\sigma]}\,
  \Bigl( R_{\alpha\beta}-\frac{R}{2(d-1)}\,g_{\alpha\beta}\Bigr)\,,
\label{Weyl_tensor2}
\end{align}
can be rewritten as a sum of Weyl-covariantized curvature tensors 
with the use of Eqs.~\eqref{Weyl_Riemann_A_F}, \eqref{Weyl_Ricci_A_F} 
and \eqref{Weyl_R_A_F}:
\begin{align}
 C_{\mu\nu\lambda\sigma} = \cR_{\mu\nu\lambda\sigma}
  -\cF_{\mu\nu}\,g_{\lambda\sigma}
  +\frac{4}{d-2}\,\,\delta^\alpha_{[\mu}g_{\nu][\lambda}\delta^\beta_{\sigma]}\,
  \Bigl( \cR_{\alpha\beta}-\frac{\cR}{2(d-1)}\,g_{\alpha\beta}
  +\cF_{\alpha\beta}\Bigr)\,.
\label{Weyl_tensor_conformal2}
\end{align}
This exhibits that $C_{\mu\nu\lambda\sigma}$ 
is a conformal tensor of weight $-2$. 
Since the tensor $L^{\lambda\sigma}_{\alpha\mu\nu\beta}\equiv
4\,\delta^\lambda_{[\alpha}g_{\mu][\nu}\delta^\sigma_{\beta]}$ 
satisfies 
$u^\alpha L^{\lambda\sigma}_{\alpha\langle\mu\nu\rangle\beta}\,u^\beta
=-\,\delta^\lambda_{\langle\mu}\,\delta^\sigma_{\nu\rangle}$\,,
we have
\footnote{
Note that 
$u^\alpha\,C_{\alpha\mu\nu\beta}\,u^\beta \,
= u^\alpha\,C_{\alpha\,\langle\mu\nu\rangle\,\beta}\,u^\beta$ 
since $C_{\alpha\mu\nu\beta}=C_{\beta\nu\mu\alpha}$ 
and $u^\alpha u^\mu C_{\alpha\mu\nu\beta} = 0 =
C_{\alpha\mu\nu\beta} u^\nu u^\beta$\,. 
}
\begin{align}
 &u^\alpha\,C_{\alpha\mu\nu\beta}\,u^\beta \,
  \bigl(\,= u^\alpha\,C_{\alpha\,\langle\mu\nu\rangle\,\beta}\,u^\beta \,\bigr)
  =u^\alpha\,\cR_{\alpha\,\langle\mu\nu\rangle\,\beta}\,u^\beta
  -\frac{1}{d-2}\,\cR_{\langle\mu\nu\rangle}\,.
\label{uCu_uRu2}
\end{align}

Multiplying Eqs.~\eqref{u_cR_u_symm} and \eqref{u_cR_u_anti} 
by $\sigma^{\mu\nu}$ and $\omega^{\mu\nu}$, respectively, 
and using Eq.~\eqref{uCu_uRu2}, 
we obtain the following formulas: 
\begin{align}
 \sigma^{\mu\nu}\, u^\alpha \, \cR_{\alpha\langle\mu\nu\rangle\beta}\, u^\beta 
 &= \sigma^{\mu\nu}\, u^\alpha\, C_{\alpha\mu\nu\beta} \, u^\beta 
  + \frac{1}{d-2}\, \cR^{\langle\mu\nu\rangle} \sigma_{\mu\nu} \nn\\
 &= \tr(\sigma^3) + \tr(\sigma\, \omega^2) + \sigma^{\mu\nu}\, 
  \cD_u \sigma_{\mu\nu}  \,,
\label{sigma_u_cR_u}\\
 \frac{1}{2}\, \cF^{\mu\nu} \omega_{\mu\nu} 
 &= -2 \tr(\sigma\, \omega^2) + \omega^{\mu\nu}\, \cD_u \omega_{\mu\nu} \,.
\label{omega_cF}
\end{align}
We can also show
\begin{align}
 \cD_\mu \cD_\nu \omega^{\mu\nu} 
 &= \frac{1}{2}\, [\cD_\mu, \cD_\nu]\, \omega^{\mu\nu}
  = -\,\cR_{[\mu\nu]}\, \omega^{\mu\nu}
  +\frac{3}{2}\,\,\cF_{\mu\nu}\,\omega^{\mu\nu} \nn\\
 &= -\,\frac{d-3}{2}\,\cF^{\mu\nu}\omega_{\mu\nu} \,, 
\label{DDomega}
\\
 \cD_\mu\bigl[ u_\nu \bigl(\cG^{\nu\mu}-\cF^{\nu\mu} \bigr) \bigr] 
  &= (\cD_\mu u_\nu)\,\Bigl(\cG^{(\mu\nu)} - \frac{d-2}{2}\,\cF^{\mu\nu}\Bigr) 
  = \cG^{\langle\mu\nu\rangle}\sigma_{\mu\nu}
  - \frac{d-2}{2}\,\cF^{\mu\nu} \omega_{\mu\nu}\nn\\
 &= \cR^{\langle\mu\nu\rangle}\sigma_{\mu\nu}
  - \frac{d-2}{2}\,\cF^{\mu\nu} \omega_{\mu\nu}\,, 
\label{DuGF}
\\
 \cD_\mu \cD_\nu \sigma^{\mu\nu} 
  &= \cD_\mu \cD_\nu \cD^\mu u^\nu 
  - \cD_\mu \cD_\nu \omega^{\mu\nu} 
  = \cD^\mu [\cD_\nu, \cD_\mu]\, u^\nu
  + \frac{d-3}{2}\,\cF^{\mu\nu}\omega_{\mu\nu} \nn\\
 &= \cD^\mu \bigl[(\cR_{\mu\nu} - \cF_{\mu\nu})\,u^\nu\bigr] 
  + \frac{d-3}{2}\,\cF^{\mu\nu}\omega_{\mu\nu} \nn\\
 &= \cD^\mu \bigl[u^\nu\bigl(\cG_{\nu\mu} - \cF_{\nu\mu}\bigr)\bigr] 
  +\frac{1}{2}\,\cD_u \cR 
  + (d-2) \,\cD^\mu \bigl[\cF_{\mu\nu}\,u^\nu\bigr] 
  + \frac{d-3}{2}\,\cF^{\mu\nu}\omega_{\mu\nu} \nn\\
 &= \cR^{\langle\mu\nu\rangle}\sigma_{\mu\nu} 
  + \frac{1}{2}\,\cD_u \cR 
  - \frac{1}{2}\,\cF^{\mu\nu}\omega_{\mu\nu}
  + (d-2) \,\cD^\mu \bigl[\cF_{\mu\nu}\,u^\nu\bigr] \,.
\label{DDsigma}
\end{align}

We can show that the spatial vector $\cF_{\mu\nu}\,u^\nu$ vanishes 
up to third-order derivatives:%
\footnote{
We denote terms of $n^{\rm th}$ and higher order derivatives by $O(\epsilon^n)$\,.
} 
\begin{align}
 \cF_{\mu\nu}\,u^\nu 
  =- \bigl(\sigma_{\mu\nu}+\omega_{\mu\nu}\bigr)\, a^\nu 
   - h_{\mu\nu} \,\nabla_u a^\nu
   + \frac{1}{d-1}\, h_{\mu}^{~\nu}\,\partial_\nu \vartheta 
  = O(\epsilon^3) \,.
\label{third-order}
\end{align}
In fact, from Eqs.~\eqref{cons_long} and \eqref{cons_trans}, we find 
\begin{align}
 \frac{1}{e}\,\cD_u e=\nabla_u \ln e
  +\frac{d}{d-1}\,\vartheta=O(\epsilon^2)\,,\qquad
  a^\mu=-\,\frac{1}{d}\,h^{\mu\alpha}\,\partial_\alpha \ln e+O(\epsilon^2)\,,
\label{acceleration_approx}
\end{align}
and thus have
\begin{align}
 h_{\mu\nu}\,\nabla_u a^\nu
  &=-\,\frac{1}{d}\,h_{\mu\nu}\,\nabla_u \Bigl(h^{\nu\alpha}\,\partial_\alpha \ln e\Bigr)
  +O(\epsilon^3)
\nn\\
 &=-\,\frac{1}{d}\,h_{\mu\nu}\,(\nabla_u h^{\nu\alpha})\,\partial_\alpha \ln e
  -\frac{1}{d}\,h_{\mu}^{~\alpha}\,\nabla_u \partial_\alpha\ln e
  +O(\epsilon^3)
\nn\\
 &=-\,\frac{1}{d}\,h_{\mu\nu}\,\bigl(\nabla_u (u^\nu u^\alpha)\bigr)\,\partial_\alpha \ln e
  -\frac{1}{d}\,h_{\mu}^{~\alpha}\,\nabla_\alpha\nabla_u \ln e
  +\frac{1}{d}\,h_{\mu}^{~\alpha}\,(\nabla_\alpha u^\beta)\,\partial_\beta \ln e
  +O(\epsilon^3)
\nn\\
 &=\frac{1}{d-1}\,a_\mu\,\vartheta + \frac{1}{d-1}\,h_{\mu}^{~\alpha}\,\partial_\alpha \vartheta
  +\Bigl(\sigma_{\mu}^{~\alpha}+\omega_{\mu}^{~\alpha}
  +\frac{1}{d-1}\,\vartheta\,h_{\mu}^{~\alpha}\Bigr)\partial_\beta \ln e
  +O(\epsilon^3)
\nn\\
 &=\frac{1}{d-1}\,h_{\mu}^{~\alpha}\,\partial_\alpha \vartheta
  -(\sigma_{\mu}^{~\alpha}+\omega_{\mu}^{~\alpha})\,a_\alpha
  +O(\epsilon^3)\,.
\end{align}

Similarly, using Eq.~\eqref{acceleration_approx}, 
we can show that 
the spatial components of $\cF_{\mu\nu}$ vanish 
up to third-order derivatives:
\begin{align}
 \frac{1}{2}\,h_{\mu}^{~\alpha}\,h_{\nu}^{~\beta}\,\cF_{\alpha\beta}
  &=  h_{\mu}^{~\alpha} h_{\nu}^{~\beta} \nabla_{[\alpha} a_{\beta\,]}
     -\frac{\vartheta}{d-1}\,\omega_{\mu\nu}  
\nn\\
 &= -\,\frac{1}{d}\,h_{\mu}^{~\alpha} h_{\nu}^{~\beta}\,
  \nabla_{[\alpha} \bigl(h_{\beta\,]}^{~\lambda}\, \partial_\lambda \ln e \bigr)
  -\frac{\vartheta}{d-1}\,\omega_{\mu\nu} + O(\epsilon^3)
\nn\\
 &= -\,\frac{1}{d}\,h_{\mu}^{~\alpha} h_{\nu}^{~\beta}\,
  (\nabla_{[\alpha}u_{\beta\,]})\,\nabla_u \ln e 
  -\frac{\vartheta}{d-1}\,\omega_{\mu\nu} + O(\epsilon^3)
\nn\\
 &= O(\epsilon^3) \,,
\end{align}
from which we have
\begin{align}
 \cF^{\mu\nu}\,\omega_{\mu\nu}=O(\epsilon^4)\,.
\label{third-order2}
\end{align}

\baselineskip=0.85\normalbaselineskip



\begin{thebibliography}{9}
\setlength{\itemsep}{-2pt}

\bibitem{LL_elasticity}
  L.~D.~Landau and E.~M.~Lifshitz, 
  ``Theory of Elasticity,'' 
  Butterworth-Heinemann (1986).

\bibitem{Eckart:1948}
  C.~Eckart,
  ``The Thermodynamics of Irreversible Processes. IV. The Theory of Elasticity
  and Anelasticity,''
  Phys.\ Rev.\  {\bf 73}, 373 (1948).

\bibitem{fs1}
  M.~Fukuma and Y.~Sakatani,
  ``Entropic formulation of relativistic continuum mechanics,''
  Phys.\ Rev.\  E {\bf 84}, 026315 (2011)
  [arXiv:1102.1557 [hep-th]].

\bibitem{fs2} 
  M.~Fukuma and Y.~Sakatani,
  ``Relativistic viscoelastic fluid mechanics,''
  Phys.\ Rev.\ E {\bf 84}, 026316 (2011)
  [arXiv:1104.1416 [cond-mat.stat-mech]].

\bibitem{Eckart:1940te}
  C.~Eckart,
  ``The Thermodynamics of Irreversible Processes. III. 
  Relativistic Theory of the Simple Fluid,''
  Phys.\ Rev.\  {\bf 58}, 919 (1940).

\bibitem{LL_fluid}
  L.~D.~Landau and E.~M.~Lifshitz, 
  ``Fluid Mechanics,'' 
  Butterworth-Heinemann (1987).

\bibitem{Romatschke}
  P.~Romatschke,
  ``New Developments in Relativistic Viscous Hydrodynamics,''
  Int.\ J.\ Mod.\ Phys.\  E {\bf 19}, 1 (2010)
  [arXiv:0902.3663 [hep-ph]].

\bibitem{fluid-gravity_review}
  M.~Rangamani,
  ``Gravity and Hydrodynamics: Lectures on the fluid-gravity correspondence,''
  Class.\ Quant.\ Grav.\  {\bf 26}, 224003 (2009)  [arXiv:0905.4352 [hep-th]].

\bibitem{Baier:2007ix}
  R.~Baier, P.~Romatschke, D.~T.~Son, A.~O.~Starinets and M.~A.~Stephanov,
  ``Relativistic viscous hydrodynamics, conformal invariance, and holography,''
  JHEP {\bf 0804}, 100 (2008)
  [arXiv:0712.2451 [hep-th]].

\bibitem{Loganayagam:2008is} 
  R.~Loganayagam,
  ``Entropy Current in Conformal Hydrodynamics,''
  JHEP {\bf 0805}, 087 (2008)
  [arXiv:0801.3701 [hep-th]].

\bibitem{Geroch:1990bw} 
  R.~P.~Geroch and L.~Lindblom,
  ``Dissipative relativistic fluid theories of divergence type,''
  Phys.\ Rev.\ D {\bf 41}, 1855 (1990).

\bibitem{PeraltaRamos:2009kg} 
  J.~Peralta-Ramos and E.~Calzetta,
  ``Divergence-type nonlinear conformal hydrodynamics,''
  Phys.\ Rev.\ D {\bf 80}, 126002 (2009)  [arXiv:0908.2646 [hep-ph]].

\bibitem{Mueller}
  I.~M\"uller,
  ``Zum Paradoxon der W\"armeleitungstheorie,''
  Z.\ Phys.\ {\bf 198}, 329 (1967).

\bibitem{Israel:1976tn}
  W.~Israel,
  ``Nonstationary irreversible thermodynamics: A Causal relativistic theory,''
  Annals Phys.\  {\bf 100}, 310 (1976).

\bibitem{Israel:1979wp}
  W.~Israel and J.~M.~Stewart,
  ``Transient relativistic thermodynamics and kinetic theory,''
  Annals Phys.\  {\bf 118}, 341 (1979).

\bibitem{Bhattacharyya:2008xc}
  S.~Bhattacharyya, V.~E.~Hubeny, R.~Loganayagam, G.~Mandal,
S.~Minwalla, T.~Morita, M.~Rangamani and H.~S.~Reall,
  ``Local Fluid Dynamical Entropy from Gravity,''
  JHEP {\bf 0806}, 055 (2008)
  [arXiv:0803.2526 [hep-th]].

\bibitem{Bhattacharyya:2008mz}
  S.~Bhattacharyya, R.~Loganayagam, I.~Mandal, S.~Minwalla and A.~Sharma,
  ``Conformal Nonlinear Fluid Dynamics from Gravity in Arbitrary Dimensions,''
  JHEP {\bf 0812}, 116 (2008)
  [arXiv:0809.4272 [hep-th]].

\bibitem{Dirac:1973gk} 
  P.~A.~M.~Dirac,
  ``Long range forces and broken symmetries,''
  Proc.\ Roy.\ Soc.\ Lond.\ A {\bf 333}, 403 (1973).

\bibitem{LL_statphys}
  L.~D.~Landau and E.~M.~Lifshitz, 
  ``Statistical Physics, Part I,'' 
  Butterworth-Heinemann (1984).

\bibitem{Azeyanagi:2009zd}
  T.~Azeyanagi, M.~Fukuma, H.~Kawai and K.~Yoshida,
  ``Universal description of viscoelasticity 
  with foliation preserving diffeomorphisms,''
  Phys.\ Lett.\  B {\bf 681}, 290 (2009)
  [arXiv:0907.0656 [hep-th]].

\bibitem{Azeyanagi:2010ab}
  T.~Azeyanagi, M.~Fukuma, H.~Kawai and K.~Yoshida,
  ``Universal description of viscoelasticity 
  with foliation preserving diffeomorphisms,'' 
  to appear in the proceedings of \emph{Quantum Theory and Symmetries 6}
  [arXiv:1004.3899 [hep-th]].

\bibitem{fs4} 
  M.~Fukuma and Y.~Sakatani,
  work in progress.

\end{thebibliography}
\end{document}